\documentclass[conference]{IEEEtran}
\IEEEoverridecommandlockouts

\usepackage{hyperref}
\usepackage{cite}
\usepackage{amsmath,amssymb,amsfonts}
\usepackage{algorithmic}
\usepackage{graphicx}
\usepackage{textcomp}
\usepackage{xcolor}
\usepackage{subcaption}
\usepackage{tabularx}
\usepackage{multicol}
\usepackage[export]{adjustbox}
\usepackage[noend,linesnumbered,ruled]{algorithm2e}
\SetAlFnt{\small}
\def\BibTeX{{\rm B\kern-.05em{\sc i\kern-.025em b}\kern-.08em
    T\kern-.1667em\lower.7ex\hbox{E}\kern-.125emX}}

\SetKwFor{ForPar}{for}{do in parallel}{end for}
\SetKw{KwAnd}{and}
\SetKw{KwAlloc}{allocate}

\SetAlFnt{\footnotesize}
\SetAlCapFnt{\small}
\SetAlCapNameFnt{\small}

\newcolumntype{L}[1]{>{\raggedright\let\newline\\\arraybackslash\hspace{0pt}}m{#1}}
\newcolumntype{C}[1]{>{\centering\let\newline\\\arraybackslash\hspace{0pt}}m{#1}}
\newcolumntype{R}[1]{>{\raggedleft\let\newline\\\arraybackslash\hspace{0pt}}m{#1}}

\newcommand{\refReviewW}[2]{\noindent\textbf{\hyperref[R#1O#2]{R#1O#2}}}
\newcommand{\refReviewD}[2]{\noindent\textbf{\hyperref[R#1D#2]{R#1D#2}}}

\definecolor{darkyellow}{RGB}{255,168,67}
\definecolor{darkgreen}{RGB}{85,130,54}
\definecolor{lightgray}{RGB}{192,192,192}
\definecolor{purple}{RGB}{148,55,255}
\definecolor{orange}{RGB}{255,147,0}
\newcommand{\smalltt}[1]{\texttt{\small #1}}

\newcommand{\blue}[1]{\textcolor{black}{#1}}

\newcommand{\green}[1]{\textcolor{darkgreen}{#1}}

\newcommand{\rx}{\textbf{RX}}
\newcommand{\cgrx}{\textbf{cgRX}}
\newcommand{\cgrxu}{\textbf{cgRXu}}
\newcommand{\bp}{\textbf{B+}}
\newcommand{\sa}{\textbf{SA}}
\newcommand{\hash}{\textbf{HT}}
\newcommand{\rtscan}{\textbf{RTScan (RTc1)}}
\newcommand{\scan}{\textbf{FullScan}}

\newcommand{\representative}[1]{$\blacktriangle_{#1}$}
\newcommand{\key}[1]{\textcolor{lightgray}{$\blacktriangle$}$_{#1}$}
\newcommand{\marker}[1]{$\triangle_{#1}$}
\newcommand{\purpletriangle}[1]{\textcolor{purple}{$\blacktriangle_{#1}$}}

\newcommand{\greentriangle}[1]{\green{$\blacktriangle_{#1}$}}

\definecolor{QueryBlue}{RGB}{68, 114, 196}

\definecolor{FirstRay}{RGB}{48, 110, 186}
\definecolor{SecondRay}{RGB}{234, 51, 36}
\definecolor{ThirdRay}{RGB}{79, 174, 91}
\definecolor{FourthRay}{RGB}{145, 144, 44}
\definecolor{FifthRay}{RGB}{223, 130, 68}
\definecolor{SixthRay}{RGB}{104, 51, 154}

\definecolor{Bucket0}{RGB}{192, 0, 0}
\definecolor{Bucket1}{RGB}{36, 90, 140}
\definecolor{Bucket2}{RGB}{85, 130, 54}

\begin{document}

\title{More Bang For Your Buck(et): Fast and Space-efficient Hardware-accelerated Coarse-granular Indexing on GPUs}

\author{\IEEEauthorblockN{Justus Henneberg}
\IEEEauthorblockA{\textit{Johannes Gutenberg University}\\
Mainz, Germany \\
henneberg@uni-mainz.de}
\and
\IEEEauthorblockN{Felix Schuhknecht}
\IEEEauthorblockA{\textit{Johannes Gutenberg University}\\
Mainz, Germany \\
schuhknecht@uni-mainz.de}
\and
\IEEEauthorblockN{Rosina Kharal}
\IEEEauthorblockA{\textit{University of Waterloo}\\
Waterloo, Canada \\
rosina.kharal@uwaterloo.ca}
\and
\IEEEauthorblockN{Trevor Brown}
\IEEEauthorblockA{\textit{University of Waterloo}\\
Waterloo, Canada \\
trevor.brown@uwaterloo.ca}
}

\twocolumn

\maketitle

\begin{abstract}
In recent work, it has been shown that NVIDIA's raytracing cores on RTX video cards can be exploited to realize hardware-accelerated lookups for GPU-resident database indexes.
This is done by materializing all keys as triangles in a 3D scene.
Lookups are performed by firing rays into the scene and utilizing the built-in index structure to detect collisions with triangles in a hardware-accelerated fashion. 
While this approach, called \textbf{R}TInde\textbf{X} (or \textbf{RX} for short), is indeed promising, it currently suffers from three limitations:
(1)~significant memory overhead per key,
(2)~slow range lookups,
and (3)~poor updateability.
In this work, we show that all three problems can be tackled by a single design change:
Generalizing \textbf{RX} to become a \textit{coarse-granular} index~\cgrx{}, which no longer indexes individual keys, but key buckets. 
We show that representing buckets in 3D space such that the lookup of a key is performed both correctly and efficiently is highly nontrivial and requires a careful orchestration of positioning triangles and firing rays in a specific sequence. 
Our experimental evaluation shows that \cgrx{} offers the most bang for the buck(et) by providing a up to $6.9\times$ higher ratio of throughput to memory footprint than comparable baselines (that support range lookups).
At the same time, \cgrx{} improves the range-lookup performance over \rx{} by up to~$15\times$ and offers practical updatability that is up to $5.6\times$ faster than rebuilding from scratch.
\end{abstract}

\begin{IEEEkeywords}
Database indexing, GPUs, raytracing
\end{IEEEkeywords}

\section{Introduction}
\label{sec:introduction}

Utilizing hardware accelerators in creative ways to speed up database operations has become increasingly popular over the last few years. A good example for this trend is \textbf{R}TInde\textbf{X}~\cite{lit:rtindex} (or \rx{} for short), which is a hardware-accelerated indexing mechanism that exploits the raytracing cores present on modern NVIDIA RTX video cards. 
The core idea of \rx{} is as follows: To index a set of key-rowID pairs, each key is represented by a corresponding triangle in a 3D scene. Then, each triangle is associated with a corresponding rowID. To perform a lookup, a ray is fired through the area in the 3D scene where the triangle representing the key is expected. If it collides with a triangle, the lookup is a hit, and the corresponding rowID is retrieved. 

To perform collision detection between rays and triangles efficiently, the video card (GPU) can create and utilize a special tree-based index structure in hardware, called a \textit{bounding volume hierarchy} (BVH), which indexes all triangles in the 3D scene.
As shown in~\cite{lit:rtindex}, \rx{} has two advantages over traditional software-based GPU-resident index structures such as~\cite{lit:bptree, lit:bptree2, lit:warpcore}: (a)~The construction and querying of the BVH is \textit{built-in}. Consequently, no complex manual hand-crafting of a highly parallel GPU-resident index structure is required. (b)~Both the BVH traversal as well as the ray-intersection tests with the candidate triangles are \textit{hardware-accelerated} by the dedicated raytracing cores, speeding up the lookup process over software-based implementations in several cases~\cite{lit:rtindex}. However, unfortunately, the approach also currently faces a set of limitations.

\vspace*{-0.3cm}
\begin{figure}[h!]
  \centering
  \begin{subfigure}[b]{.32\linewidth}
    \includegraphics[width=\linewidth, trim={0.2cm 0 0 0}, clip]{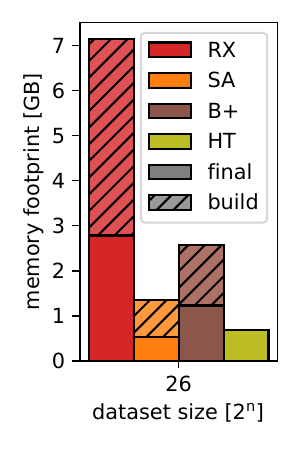}
    \caption{Memory footprint}
    \label{fig:weaknesses:footprint}
  \end{subfigure}
  \begin{subfigure}[b]{.32\linewidth}
    \includegraphics[width=\linewidth, trim={0.2cm 0 0 0}, clip]{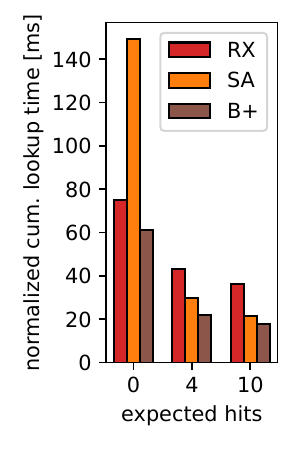}
    \caption{Range lookups}
    \label{fig:weaknesses:range_lookups}
  \end{subfigure}
  \begin{subfigure}[b]{.32\linewidth}
    \includegraphics[width=\linewidth, trim={0.2cm 0 0 0}, clip]{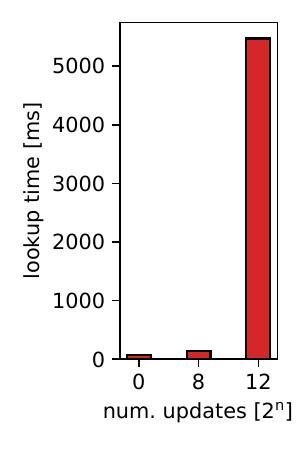}
    \caption{Updates}
    \label{fig:weaknesses:updates}
  \end{subfigure}
  \caption{Limitations of \rx{}: Memory overhead, range lookups, and lookup performance after updates~\cite{lit:rtindex}.}
    \label{fig:weaknesses}
    \vspace*{-0.2cm}
\end{figure}

First of all, the memory overhead per key of \rx{} is high since a single $8$B integer key is represented as a triangle described by nine $4$B floats, \blue{resulting in $36$B per key}. Consequently, $78\%$ of the key representation is actually overhead. In the left plot in Figure~\ref{fig:weaknesses}, which has been generated from the results of~\cite{lit:rtindex}, we can see that the traditional index structures have a significantly smaller memory footprint than \rx{}, which is mainly due to having less overhead per key. As memory is scarce on GPUs, this can be a limitation for many applications. 
Second, range lookups are currently a weakness of~\rx{}. In the middle plot in Figure~\ref{fig:weaknesses}, we can see that for all tested range sizes, the~\bp{}-tree outperforms \rx{}. The reason for this is that a range lookup in \rx{} requires a large number of ray intersection tests with candidate triangles if its selectivity is low. In contrast, a \bp{}-tree simply performs a \textit{single} tree traversal for the lower bound key and then sequentially scans the leaf level.
Third, \rx{} is very sensitive to updates. Interestingly, the problem is not the cost of performing the updates, but a severe drop in lookup performance \textit{after} the updates have been applied. The right plot in Figure~\ref{fig:weaknesses} shows this by performing a batch of lookups after applying a varying number of updates: The more updates have been applied, the more the lookup performance deteriorates, up to a slowdown of $78\times$ over no updates. The reason for this is that the BVH update procedure only scales the existing bounding volumes to reflect the updates instead of restructuring the BVH. This can heavily increase the number of intersection tests that must be carried out during lookups. 

\subsection{Challenges of Hardware-accel. Coarse-granular Indexing}

Interestingly, the aforementioned limitations can be addressed by making a single design change:
Instead of creating a \textit{fine-granular index}~\rx{}, which maps each individual key/triangle to its rowID, we propose to generalize the concept to a \textit{coarse-granular index}~\cgrx{}, which indexes groups of keys instead, \blue{but still retains hardware acceleration}. Precisely, each group is represented by a single key/triangle, which then maps to a separately stored bucket of key-rowID pairs. This design change drastically reduces the memory footprint of the structure --- for example, for a bucket size of~eight, the memory \textit{overhead} decreases from $78\%$ in \rx{} to only $36\%$ in \cgrx{}, \blue{since we only need one additional 36B triangle for every bucket of eight 8B keys}. As the number of triangles correlates with the size of the generated BVH, \cgrx{} also constructs a significantly smaller BVH than \rx{}, reducing the traversal time.
Of course, additional cost must be factored in for post-filtering a retrieved bucket for the key(s) of interest. Still, by adjusting the granularity at which keys are grouped, we can balance the cost of traversing the BVH, the cost of searching the bucket, and the overall memory footprint depending on the requirements of the environment. 

\blue{Note that while the basic principle of coarse-granular indexing is straightforward, mapping it to hardware-accelerated raytracing, which we propose in this work in form of~\cgrx{}, is highly non-trivial. As we no longer materialize \textit{all} keys in the scene, but only a \textit{single} representative key for each bucket, the central challenge is to ensure that the lookup procedure still detects all hits and misses correctly. As we will see, depending on the situation, this requires firing a sequence of up to five rays from specific positions in specific directions. Since at the same time, the performance overhead of a lookup must remain small, we will apply a set of delicate optimizations to both the triangle arrangement as well as the ray firing procedure.}

\subsection{Contributions and Structure of the Paper}

In summary, we make the following contributions: 
\textbf{(1)}~After recapping \rx{} in Section~\ref{sec:background}, we present~\cgrx{}, our new hardware-accelerated coarse-granular index for NVIDIA RTX GPUs in Section~\ref{sec:cgrx}, which supports 64-bit keys as well as point and range lookups. We first discuss the construction and lookup procedure of a \textbf{naive representation}, which speeds up the navigation through the 3D scene by introducing additional marker triangles. 
\textbf{(2)}~Based on that, we present an \textbf{optimized representation}, which avoids the materialization of additional marker triangles altogether. Instead, we turn a subset of representatives into implicit markers. This is done by (a)~moving certain representatives and (b)~introducing auxiliary representatives in the scene. The evaluation shows that the optimized representation improves both lookup performance and memory footprint for very sparse key sets. 
\textbf{(3)}~We present an extension called \cgrxu{} to support efficient batch-wise \textbf{updates} (inserts and deletes) in Section~\ref{sec:updates}. Insertions and deletions are handled by organizing buckets as a linked list of physical nodes, which are attached (or detached) on demand. By this, updates to the BVH and hence the extreme deterioration of the lookup performance is avoided.
\textbf{(4)}~As \cgrx{} provides a set of \textbf{configuration parameters}, we analyze their impact for a variety of key distributions in Section~\ref{sec:configuration}. Precisely, we analyze the impact of (a)~the optimizations, (b)~the key mapping into 3D space, as well as (c)~the bucket size.
\blue{We also test the robustness of the choice by evaluating $4560$~different indexing scenarios}.  
\textbf{(5)}~Using the best configuration(s), we perform an extensive \textbf{experimental evaluation} against a set of state-of-the-art baselines in Section~\ref{sec:exp}. The evaluation analyzes (a)~the throughput to memory footprint ratio for point lookups, (b)~the range lookup performance, (c)~the impact of batching, (d)~the impact of the hit rate, (e)~the impact of lookup skew, and (f)~the update performance.  

\section{Background}
\label{sec:background}

We start by discussing the working principle of the fine-granular index~\rx{}~\cite{lit:rtindex}, where we discuss both its construction and lookup procedure. As both \cgrx{} and \rx{} use NVIDIA's OptiX computing API~\cite{lit:optix-paper, lit:optix-homepage} to program the raytracing pipeline, the following implementation details will also be relevant for the presentation of \cgrx{} in Section~\ref{sec:cgrx}. 

\subsection{Construction of \rx{}}
\label{ssec:rx_construction}

\begin{figure}[t!]
\centering
\includegraphics[page=28, width=\columnwidth, trim={0.1cm 12cm 19.6cm 0.2cm}, clip]{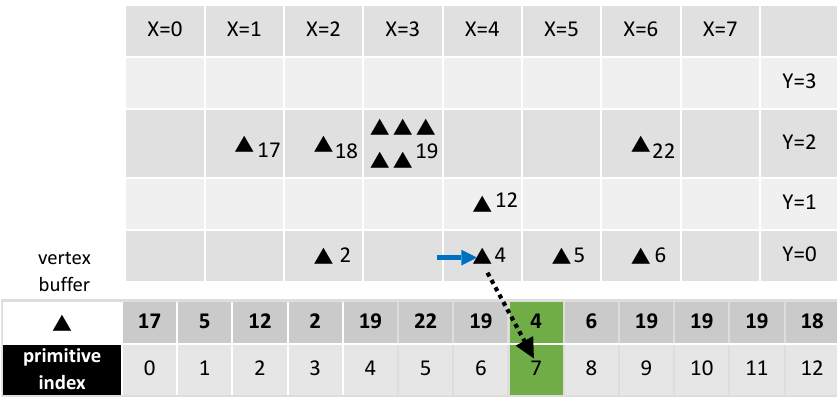}
\vspace*{-0.6cm}
\caption{Example key set and associated triangle representation for \rx{}~\cite{lit:rtindex}, followed by a lookup of key~4 returning rowID~7.}
\label{fig:rx}
\vspace*{-0.2cm}
\end{figure}

Given a set of key-rowID pairs to index, the construction happens in two steps:
In the first step, the keys are transformed into a set of corresponding triangles in the 3D scene, where for each key~$k$, \rx{} creates a single isolated triangle. Practically, this is done by writing the positions of the three corner points of each triangle one after the other into a so-called \textit{vertex buffer}. The position of each triangle in the scene is computed using a \textit{key mapping}. \cite{lit:rtindex} observed that this key mapping cannot be arbitrary, but is limited to 23 bits in each dimension to ensure correct floating-point arithmetic. Consequently, \rx{} uses a mapping where the 23 least significant bits of each key~$k$ are treated as the $x$~coordinate, the next 23 bits as the $y$~coordinate, and the 18 most significant bits as the $z$~coordinate, denoted as \mbox{$k \mapsto (k_{22:0}, k_{45:23}, k_{63:46})$}. Geometrically speaking, this key mapping arranges all triangles into \textit{rows} and \textit{planes}. 
Note that to ease visualization, in the following examples, we will use a simpler key mapping where the three last bits of the key determine the $x$~coordinate, the next two bits determine the $y$~coordinate, and the remainder determines the $z$~coordinate, i.e., \mbox{$k \mapsto (k_{2:0}, k_{4:3}, k_{63:5})$}. Apart from mapping the keys to triangles, each key~$k$ also must be associated with its rowID~$r$. This is done implicitly by materializing the triangle of~$k$ at position~$r$ in the vertex buffer. This position is called the \textit{primitive index}, and it can be queried later on.
Figure~\ref{fig:rx} shows how the generated triangles look for an example key set of $13$~keys, as well as the state of the vertex buffer. For example, key~$4$ is mapped to the triangle~\representative{4} at position~$x=4$, $y=0$, and $z=0$. This triangle is materialized in the vertex buffer at slot~$7$. This effectively associates key~$4$ with rowID~$7$.

\begin{figure}[t!]
\centering
\includegraphics[page=29, width=\columnwidth, trim={0.1cm 14.3cm 19.2cm 0.2cm}, clip]{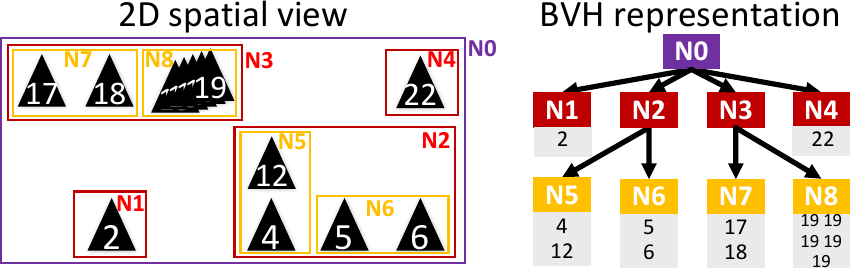}
\vspace*{-0.3cm}
\caption{Possible BVH for the scene in Figure~\ref{fig:rx}.}
\label{fig:rxbvh}
\vspace*{-0.3cm}
\end{figure}

In the second step, the aforementioned BVH is built on top of the 3D scene. 
A BVH is a tree-like index structure in which small, disjoint groups of triangles form the leaves.
Each group is then enclosed by a 3D cuboid, a so-called bounding volume.
These bounding volumes are then iteratively grouped and enclosed by larger bounding volumes until only the root bounding volume remains. 
To construct the BVH, the vertex buffer is passed to \smalltt{optixAccelBuild()}, which then indexes all individual triangles in the buffer without any further involvement of the programmer. 
Figure~\ref{fig:rxbvh} visualizes how the generated BVH could look like (in 2D) for the example key set. In the example, the triangle representing key~$4$ is enclosed by the small bounding volume~N5 first, next by the larger volume~N2, and finally by the root volume~N0.      

\subsection{Lookups in \rx{}}
\label{ssec:rx_lookups}

Using the generated BVH, NVIDIA's OptiX is able to quickly find intersections between triangles and rays via hardware acceleration.
Therefore, \rx{} maps each lookup operation to a corresponding ray-triangle intersection problem.
A ray is defined by its point of origin~$o$ and a three-dimensional direction vector~$d$.
To perform a point lookup of key $k$, one first computes the 3D position $p$ associated with $k$ using the key mapping, and then selects the ray parameters $o$ and $d$ so that the ray passes through $p$.
If a triangle exists at position $p$, the ray will intersect this triangle and retrieve the associated rowID.
To prevent a ray from extending beyond a single triangle and producing false positives, OptiX provides an option to limit a ray to a specified length.
Similarly, a range lookup~$[l,u]$ can be performed by firing one or multiple rays in parallel to the $x$-axis, starting at the position associated with the lower bound~$l$, and limiting the ray to not extend beyond the upper bound~$u$.
This way, only the triangles that are located between the given bounds will be hit, and we eventually obtain all rowIDs that are relevant to the lookup.
As is typical for GPU-resident indexes, \rx{} implements batch lookups to improve GPU utilization, where
each lookup is performed by a single thread.

\blue{Note that the execution time of a ray traversal depends on the shape of the BVH, and ultimately on the key distribution. Average case analyses for ray traversal times following several common BVH construction algorithms are surveyed in~\cite{lit:bvhtraversalcost}.}

\section{Coarse-granular RX}
\label{sec:cgrx}

Let us now discuss the specifics of~\cgrx{}.
Using a running example, we will present two different 3D scene representations that ~\cgrx{} can generate and query:
A \textbf{naive representation}, and an \textbf{optimized representation} which requires fewer rays to be cast in certain situations while \textit{also} having a lower memory footprint. 

\begin{figure}[t]
\centering
\includegraphics[page=3, width=\columnwidth, trim={0.2cm 10cm 19.6cm 0.2cm}, clip]{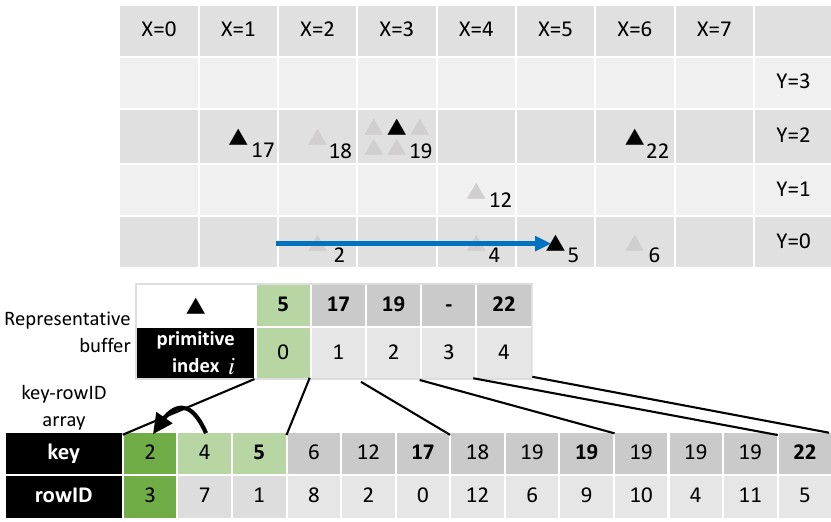}
\vspace*{-0.3cm}
\caption{Example key set and associated triangle representation, followed by a lookup of key~2 which returns rowID~3. The representative~\representative{5} of bucket~0 is located in the \textit{same} row as the searched key~2. Row markers are hidden for simplicity.}
\label{fig:cgrxTwoD_query_singlerow}
\vspace*{-0.2cm}
\end{figure}


Just like \rx{}, \cgrx{} uses a key mapping to uniquely represent each key as a triangle on an integer grid, which is exemplified in Figure~\ref{fig:cgrxTwoD_query_singlerow}.
However, in contrast to \rx{}, not \textit{all} keys are actually materialized as triangles. An array of keys and rowIDs is sorted and logically partitioned into equally-sized buckets (of size 3), and only the last key in each bucket is inserted as a \textit{bucket representative}, shown as a black triangle~\representative{k}. All remaining keys are \textit{not} materialized in the scene. We still visualize them as gray triangles~\key{k}.
As a consequence of this design, the number of triangles we need to store in the vertex buffer is greatly reduced over \rx{} which, in turn, reduces the size of the BVH.
At the same time, looking up a key~$k$ becomes more complex, since there is no guarantee that there will be a triangle at the position~$p$ associated with~$k$.
Instead, we need to search for the next bucket representative, which is always larger than or equal to~$k$.
So, the bucket representative either has to (1) be in the same row, but have a larger or equal $x$~coordinate than~$p$, or (2) be on the same plane, but have a larger $y$~coordinate than~$p$, or (3) be on a different plane and have a larger $z$~coordinate than $p$.

\noindent\textbf{Case (1)} is visualized in Figure~\ref{fig:cgrxTwoD_query_singlerow} by showing the lookup of key~$2$, where the triangle~\key{2} does not actually exist in the scene:
To locate the bucket representative, we cast a \textcolor{FirstRay}{single ray} along the positive $x$-axis, starting our ray slightly left of~\key{2}.
The ray intersects~\representative{5}, which is the first triangle in the vertex buffer, and is therefore associated with primitive index~$0$ and hence bucket~$0$.
We then search bucket~$0$ to find that key~$2$ occurs at rowID~$3$ in the original table (column).

\begin{figure}[!h]
\centering
\includegraphics[page=4, width=.95\columnwidth, trim={0.2cm 8.7cm 19.6cm 0.2cm}, clip]{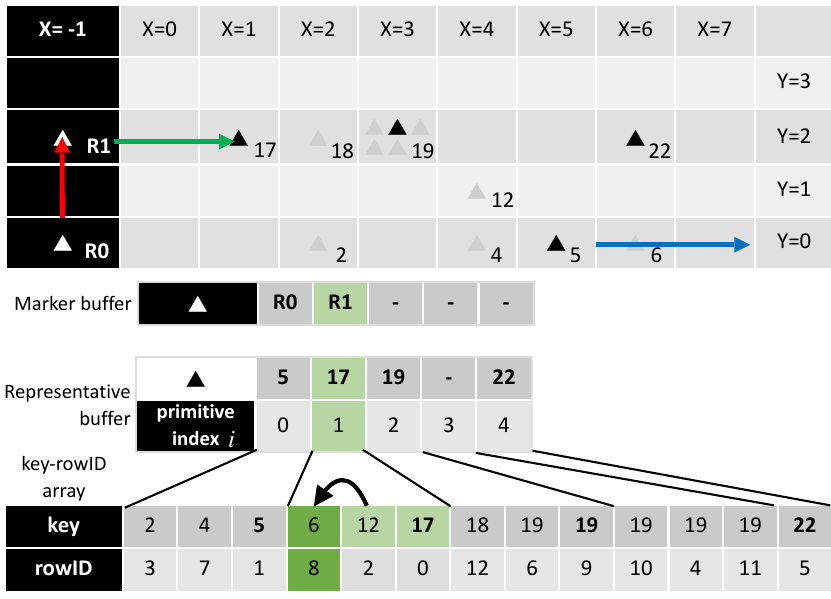}
\vspace*{-0.2cm}
\caption{Lookup of key~6 which returns rowID~8. The representative~\representative{17} of the corresponding bucket~1 is located in a \textit{different} row as the searched key~6. This row is quickly identified using marker~\marker{R1}.}
\label{fig:cgrxTwoD_query_multirow}
\end{figure}

\noindent\textbf{Case (2)}, depicted in Figure~\ref{fig:cgrxTwoD_query_multirow}, is more complex:
When attempting to find key~$6$, the \textcolor{FirstRay}{first ray} fails to hit a representative in the same row.
Thus, the next representative must be the leftmost triangle in the next populated row.
To allow efficient discovery of the next populated row, we introduce \textit{row markers}~\marker{R} into the scene:
If a row contains at least one representative, we add a triangle at $x=-1$ in the same row.
With the help of row markers, we can easily locate the next populated row by casting a \textcolor{SecondRay}{second ray} along the $y$-axis starting in the subsequent row at $x=-1$.

After intersecting~\marker{R1} at $y=2$, we cast a \textcolor{ThirdRay}{third ray} from $y=2, x=0$ to find the first representative of this row, which hits triangle~\representative{17} associated with bucket~$1$.
Note that in the figure, the $y=2$ row contains multiple triangles that could be intersected by the ray.
However, we are only interested in the leftmost intersection, i.e., the one closest to the ray origin.
This \textit{closest-hit} discovery is a fundamental operation in computer graphics, and therefore, natively supported by OptiX.

\begin{figure}[h!]
    \centering
    \includegraphics[page=27, width=.95\columnwidth, trim={0.2cm 4.5cm 15.5cm 0.2cm}, clip]{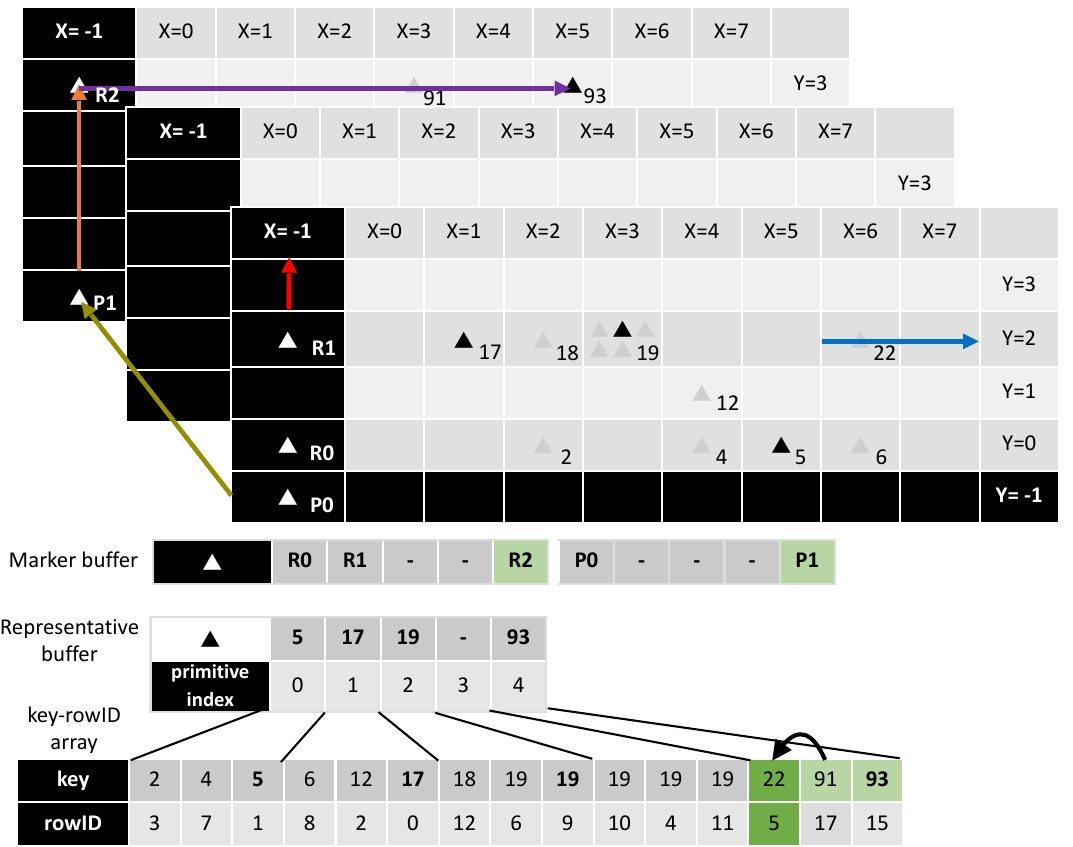}
    \vspace*{-0.2cm}
    \caption{Lookup of key~22 when the key set is spread across multiple planes. The example shows the worst case where five rays are required to perform the lookup.}
    \label{fig:cgrx_threeD_query}
\end{figure}

\noindent\textbf{Case (3)} is similar to case (2).
It is shown in Figure~\ref{fig:cgrx_threeD_query}, where a lookup of key~$22$ is performed on an extended key set. 
If the \textcolor{SecondRay}{second ray} fails to hit a row marker, there are no more populated rows on this plane.
Therefore, we now need to find the next populated plane.
Similar to what we did with rows, we also mark every populated plane with a \textit{plane marker}~\marker{P} at $x=-1$ and $y=-1$.
This allows us to cast a \textcolor{FourthRay}{ray along the $z$-axis} to discover~\marker{P1}, followed by two more rays:
\textcolor{FifthRay}{One ray} cast along the $y$-axis locates the next populated row via~\marker{R2}, and the \textcolor{SixthRay}{final ray} along the $x$-axis intersects triangle~\representative{93} associated with bucket~$4$.

Let us also discuss how to \textbf{handle duplicates}:
\blue{If the same key occurs multiple times, it can happen that the duplicates of the key span over multiple buckets. This is the case for key~$19$ in Figure~\ref{fig:cgrx_threeD_query}, which occurs five times in total and consequently spans over the buckets~$2$~and~$3$. To handle this situation, we create a representative only for the first of the two buckets. A lookup for $19$ will then find this first representative and consequently jump to the start of bucket~$2$ in the sorted key-rowID array. Scanning the bucket retrieves the first duplicate of~$19$. Scanning further will identify all remaining duplicates of $19$ across buckets. The scan stops as soon as the first key larger than $19$ is found, namely~$22$. This ensures that \textit{all} duplicates are visited.}
Similarly, we only generate a marker for the first representative in the row/plane to avoid duplicate markers.

\subsection{Implementation Details}
\label{ssec:cgrx_impl}

\textit{Construction.} 
In Algorithm~\ref{listing:cgrx_construction}, we formalize the construction procedure as pseudo-code.
We use the notation $k.x$ to refer to the bits of $k$ that are mapped to the $x$-coordinate.
These bit operations are required several times:
In lines \ref{line:rmark} and \ref{line:pmark}, we check if all representatives lie on the same plane or even in the same row, in which case we can skip the allocation and generation of plane/row markers.
The algorithm then loops over all buckets and creates a representative for each bucket, if necessary.  
In lines \ref{line:new_row} and \ref{line:new_plane}, we check whether the previous key belongs to a different row/plane, in which case the current key is the first of the current row/plane.
The mkTri$(x, y, z)$ function creates a small triangle that is centered around the point~$(x, y, z)$.
\blue{Note that for simplicity, the pseudo-code does not show the specific handling of the first iteration, where there exists no previous representative. Of course, our actual implementation handles this case correctly.}

\begin{algorithm}
\caption{Construction of the naive representation}
\label{listing:cgrx_construction}
\KwIn{keys, bucketSize}
\KwOut{reps, markers}
minRep $\gets$ keys[bucketSize - 1], maxRep $\gets$ keys[len(keys) - 1] \\
multiLine $\gets$ minRep.yz != maxRep.yz  \\ \label{line:rmark}
multiPlane $\gets$ minRep.z != maxRep.z  \\ \label{line:pmark}
numBuckets $\gets$ ceil(len(keys) / bucketSize)\\ \label{line:num_buckets}
\KwAlloc reps[numBuckets] \\
\KwAlloc markers[(multiLine + multiPlane) $\cdot$ numBuckets] \\

\ForPar{\textnormal{bucketID} $\gets0$ \KwTo \textnormal{numBuckets - 1}}{ \label{line:loop}
    repIdx $\gets$ min((bucketID + 1) $\cdot$ bucketSize, len(keys)) - 1\\ \label{line:idx}
    rep $\gets$ keys[repIdx]\\ \label{line:key}
    \blue{prevRep $\gets$ keys[repIdx - bucketSize]}\\ \label{line:prev_key}

    \uIf{\textnormal{rep != prevRep}}{ \label{line:key_vs_prev_key}
        reps[bucketID] = mkTri(rep.$x$, rep.$y$, rep.$z$)\\ \label{line:add_rep}
    }
    \uIf{\textnormal{multiLine \KwAnd rep.$yz$ != prevRep.$yz$}}{ \label{line:new_row}
        markers[bucketID] = mkTri(-1, rep.$y$, rep.$z$)\\ \label{line:add_row_marker}
    }
    \uIf{\textnormal{multiPlane \KwAnd rep.$z$ != prevRep.$z$}}{ \label{line:new_plane}
        markers[bucketID + numBuckets] = mkTri(-1, -1, rep.$z$)\\ \label{line:add_plane_marker}
    }
}
\Return reps, markers
\label{line:return_buffers_min_max}
\end{algorithm}

\textit{Lookups.} Algorithm~\ref{listing:cgrx_lookup} shows the code responsible for obtaining the bucketID for a given key.
The $x$Cast$(x, y, z)$ function internally delegates to OptiX to cast a ray with direction $(1, 0, 0)$ originating at $(x, y, z)$.
Its return value stores whether a triangle was intersected by the ray and, if so, exposes the primitive index as well as the coordinates of the intersection point.
$y$Cast and $z$Cast are defined analogously.
The \textcolor{SecondRay}{first $y$-axis ray} and the \textcolor{FourthRay}{$z$-axis ray} have to start in the next row/plane, so we offset their origins by 1 in the appropriate directions.
Note that Algorithm~\ref{listing:cgrx_lookup} can also be used to answer range lookups of the form~$[l, u]$:
We utilize the raytracing approach to find the representative for $l$, which leads us to the first bucket containing a value larger than or equal to $l$.
From there, we linearly scan the key-rowID array until we hit the first key larger than $u$. If the lower bound~$l$ is larger than the largest key, we can safely report an empty result. \blue{\cgrx{} always performs this scan by invoking a separate CUDA kernel to spawn a group of 16 threads per lookup, which allows loading (and potentially aggregating) neighboring entries very efficiently.}


\textit{Markers.} 
\blue{Whenever all representatives share the same plane or row, we can skip the generation of some or all markers.
We do so by retrieving the smallest representative and the largest representative of the dataset and checking whether they span across multiple rows (line~\ref{line:key_range_check_first}) and multiple planes (line~\ref{line:key_range_check}) by inspecting their coordinates. During construction, we then add row/plane markers only if they are actually required (lines~\ref{line:new_row}-\ref{line:add_plane_marker}).}
The effect of this optimization is shown in Figures~\ref{fig:cgrxTwoD_query_singlerow} and~\ref{fig:cgrxTwoD_query_multirow}, which only contain row markers.
Even so, in the worst-case, the use of markers can still inflate the size of the BVH by 3x, which is a problem that we will address in the optimized representation.

\begin{algorithm}
\DontPrintSemicolon
\caption{Point lookup in the naive representation}
\label{listing:cgrx_lookup}
\KwIn{key}
\KwOut{bucketID or MISS}
\blue{minRep $\gets$ keys[bucketSize - 1], maxRep $\gets$ keys[len(keys) - 1]} \\
\lIf{\textnormal{k \textless{} minRep}}{
    \Return 0 \label{line:key_range_check_first}
}
\lIf{\textnormal{k \textgreater{} maxRep}}{
    \Return MISS \label{line:key_range_check}
}
\textcolor{FirstRay}{sameRowHit $\gets$ $x$Cast(key.$x$, key.$y$, key.$z$)} \\ \label{line:first_ray}
\lIf{\textnormal{\textcolor{FirstRay}{sameRowHit}}}{ \label{line:first_ray_hits}
    \Return \textcolor{FirstRay}{sameRowHit}.primitiveIndex
}
\textcolor{SecondRay}{nextRowHit $\gets$ $y$Cast(-1, key.$y$ + $1$, key.$z$)} \\ \label{line:second_ray}
\uIf{\textnormal{\textcolor{SecondRay}{nextRowHit}}}{ \label{line:second_ray_hits}
    \textcolor{ThirdRay}{sameRowHit $\gets$ $x$Cast($0$, nextRowHit.$y$, key.$z$)}\\ \label{line:third_ray}
    \Return \textcolor{ThirdRay}{sameRowHit}.primitiveIndex
}
\textcolor{FourthRay}{nextPlaneHit $\gets$ $z$Cast(-1, -1, key.$z$ + $1$)} \\ \label{line:fourth_ray}
\label{line:fourth_ray_hits}
\textcolor{FifthRay}{nextRowHit $\gets$ $y$Cast(-1, $0$, nextPlaneHit.$z$)} \\  \label{line:fifth_ray}
\textcolor{SixthRay}{sameRowHit $\gets$ $x$Cast($0$, nextRowHit.$y$, nextPlaneHit.$z$)}\\ \label{line:sixth_ray}
\Return \textcolor{SixthRay}{sameRowHit}.primitiveIndex
\end{algorithm}

\textit{Bucket Search.} To search a bucket, \cgrx{} supports both linear search and binary search on buckets in column layout as well as in row layout. However, our experimental evaluation showed that both for small buckets of~$4$ entries and very large buckets of $65{,}536$ entries, binary search on a row layout performs best, so we use this combination for the remainder of the paper.


\subsection{Optimized Representation}
\label{ssec:cgrx_opt}

Depending on the key distribution, it is possible that a single bucket spans multiple rows or even multiple planes.
This is the reason why we have to potentially fire multiple rays to locate a representative.  
As firing more rays makes the lookup more expensive, in the following, we propose an optimized representation which addresses this problem while potentially decreasing the memory footprint. 
The high-level idea is based on two modifications which we are allowed to perform in the scene without harming correctness: 
(1)~\label{enum:opt_observation_replace} Let $r$ be a representative and let $k$ be the next key after $r$.
Observe that we can 
replace $r$ by another representative~$r'$ as long as $r < r' < k$, even if $r'$ is not a key itself. 
In other words, we can \textit{move} a representative as long as it does not collide with the next key.
(2)~\label{enum:opt_observation_insert} Let $r$ and $r'$ be two adjacent representatives with $r < r'$.
Then we are allowed to \textit{insert} a new representative~$r''$ between them so that $r < r'' < r'$.
Since $r''$ falls into the bucket represented by $r'$, we need to associate $r''$ with the same bucketID as $r'$.
Using these rules, we are able to ensure that each populated row ends with a representative in the last slot (at $x_{max}$),
either by moving an existing representative there (1), or by inserting a new one (2).
Consequently, when starting a lookup in a populated row, we never have to fire more than a single ray as we will always find a representative there, therefore requiring less rays along the $y$-axis.
Analogously, we can place a representative in the last slot of each populated plane to reduce the number of rays along the $z$-axis.
At the same time, these newly inserted representatives can also serve as row/plane markers since they are always located at $x=x_{max}$ or $y=y_{max}$, as long as we change the respective offsets for yCast and zCast in Algorithm~\ref{listing:cgrx_lookup}.

\begin{algorithm}
\DontPrintSemicolon
\caption{Construction of the optimized representation}
\label{listing:cgrx_opt_construction}
\KwIn{keys, bucketSize}
\KwOut{reps}
minRep $\gets$ keys[bucketSize - 1], maxRep $\gets$ keys[len(keys) - 1] \\
multiLine $\gets$ minRep.yz != maxRep.yz  \\ \label{line:opt:multi_line}
multiPlane $\gets$ minRep.z != maxRep.z  \\ \label{line:opt:multi_plane}
numB $\gets$ ceil(len(keys) / bucketSize)\\ \label{line:opt:num_buckets}
\KwAlloc reps[(1 + multiLine + multiPlane) $\cdot$ numB]\\ \label{line:opt:alloc}

\ForPar{\textnormal{bucketID} $\gets0$ \KwTo \textnormal{numB - 1}}{ \label{line:opt:loop}
    repIdx $\gets$ min((bucketID + 1) $\cdot$ bucketSize, len(keys)) - 1\\ \label{line:opt:idx}
    rep $\gets$ keys[repIdx]\\ \label{line:opt:rep}
    nextKey $\gets$ keys[repIdx + 1]\\ \label{line:opt:next_key}
    movable $\gets$ nextKey.$yz$ != rep.$yz$\\ \label{line:opt:set_do_replace}
    \blue{prevRep $\gets$ keys[repIdx - bucketSize]}\\ \label{line:opt:prev_rep}
    \blue{nextRep $\gets$ keys[repIdx + bucketSize]}\\ \label{line:opt:next_rep}

    needsRep $\gets$ rep != prevRep \textbf{or} (movable \textbf{and} rep.$x$ != $x_{max}$) \\ \label{line:opt:insert_rep_condition}
    needsRowMark $\gets$ !movable \textbf{and} rep.$yz$ != nextRep.$yz$ \\ \label{line:opt:next_rep_in_different_row}
    needsPlaneMark $\gets$ rep.$y$ != $y_{max}$ \textbf{and} rep.$z$ != nextRep.$z$ \\ \label{line:opt:next_rep_in_different_plane}
    \uIf{\textnormal{needsRep}}{ 
        \textnormal{$x \gets$ \textbf{if} movable \textbf{then} $x_{max}$ \textbf{else } rep.$x$}\\  \label{line:opt:x_position_rep}
        doFlip $\gets$ movable \textbf{and} prevRep.$yz$ != rep.$yz$ \\ \label{line:opt:do_flip}
        reps[bucketID] $\gets$ mkTri($x$, rep.$y$, rep.$z$, doFlip)\\ \label{line:opt:materialize_rep}
    }
    \uIf{\textnormal{multiLine \textbf{and} needsRowMark}}{ \label{line:opt:row_marker_condition}
        reps[bucketID + numB] $\gets$ mkTri($x_{max}$, rep.$y$, rep.$z$) \label{line:opt:materialize_new_rep_as_row_marker}
    }
    \uIf{\textnormal{multiPlane \textbf{and} needsPlaneMark}}{ \label{line:opt:plane_marker_condition}
        reps[bucketID + 2 $\cdot$ numB] $\gets$ mkTri($x_{max}$, $y_{max}$, rep.$z$) \label{line:opt:materialize_new_rep_as_plane_marker}
    }
}
\Return reps
\end{algorithm}

\textit{Construction.} Algorithm~\ref{listing:cgrx_opt_construction} shows the pseudo-code for constructing this optimized representation.
The code copies the single-row/single-plane optimization from the naive variant (lines~\ref{line:opt:multi_line} and~\ref{line:opt:multi_plane}), but instead of allocating a separate marker buffer, it reserves additional space in the representative buffer (line~\ref{line:opt:alloc}), as the optimized variant does not differentiate between representatives and markers.

For each bucket, we need to check several conditions before placing the triangles.
Following the observations listed above, a representative can be moved to the end of the row if the next key is not in the same row (line~\ref{line:opt:set_do_replace}).
This creates a special case for handling duplicate representatives:
In the naive variant, we skipped insertion of all but the first representative to ensure that no two representatives exist at the same coordinates.
In this variant, it is still possible to also insert the last representative of a duplicate group \textit{if} it can be moved away from its initial position to the end of the row (line~\ref{line:opt:insert_rep_condition}).
If a representative is the last in its row and it cannot be moved, we have to explicitly insert a new representative at the end of the row (line~\ref{line:opt:next_rep_in_different_row}).
For bucket $b$, this triangle is placed in slot $b$ + numBuckets of the vertex buffer.
Similarly, the bucket with the last representative on a plane generates an additional plane marker at $x=x_{max}$, $y=y_{max}$ in slot $b$ + 2 $\cdot$ numBuckets.
If this last representative happens to be located in the last row of a plane ($y=y_{max}$), we can skip its generation, since it coincides with the row marker (line~\ref{line:opt:next_rep_in_different_plane}).

In line~\ref{line:opt:do_flip}, we perform a further optimization called \textit{triangle flipping}.
It applies whenever a representative can be moved to the end of the row while also being the only representative in this row.
In this case, any ray being fired in the corresponding row will \textit{always} hit this representative.
Therefore, we do not need to fire this ray at all.
To inform the lookup procedure of this fact, we ``flip'' the triangle by inverting the order in which the corner points are stored in the buffer (the winding order) from clockwise to counter-clockwise.
This way, any ray fired along the $y$-axis will recognize the hit as a \textit{back-side hit} as opposed to a \textit{front-side hit}, and can react accordingly.

\begin{figure}[!h]
\centering
\includegraphics[page=24, width=.95\columnwidth, trim={0cm 9.6cm 19.5cm 0}, clip]{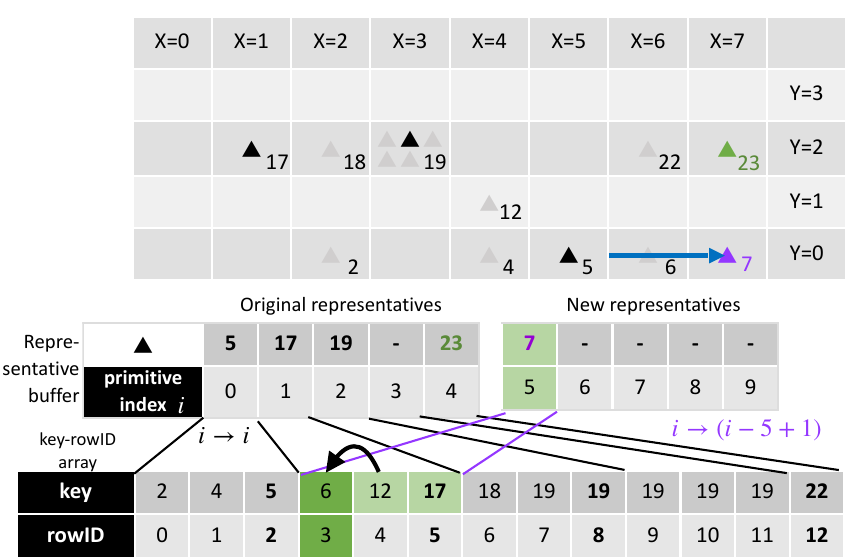}
\vspace*{-0.2cm}
\caption{Visualization of the optimized representation.}
\label{fig:cgrx_optimized}
\end{figure}

Figure~\ref{fig:cgrx_optimized} shows the optimized representation for our example key set.
In comparison to the naive representation in Figure~\ref{fig:cgrxTwoD_query_singlerow}, we can see three differences:
One representative is newly inserted as a marker (\purpletriangle{7}) while another representative is moved (\greentriangle{23} replaces the representative~\representative{22} to its left) to serve as another marker.
In exchange, no explicit markers are present anymore at $x=-1$.
Let us revisit the special cases from Algorithm~\ref{listing:cgrx_opt_construction}:
For the first bucket with bucketID~$0$, the representative~\representative{5} must be materialized, as it is not a duplicate.
However, we cannot materialize it at the end of the row, as its next key~\key{6} is located in the same row.
Still, as~\representative{5} is the last representative of the row, it is responsible for creating a new representative~\purpletriangle{7} which serves as the row marker.
Key~$19$ in row~$y=2$ appears multiple times, but since none of its instances are the last key in the row, there is no difference in how the representatives are placed compared to the naive algorithm.
Instead, we move representative~\representative{22}, which happens to be the last key in the row, to the end of the row, producing~\greentriangle{23}.
Finally, we do not insert any plane markers, since all keys reside on a single plane.
If we had to generate a plane marker, it would be located in the very last slot of the plane at $x=7$ and $y=3$.
A lookup of key~$6$~(hit) is now answered by firing a \textcolor{FirstRay}{\textit{single} ray} (hitting~\purpletriangle{7}), instead of \textit{three} rays.
Note that the primitive index $i=5$ of~\purpletriangle{7} is greater than the number of buckets, since~\purpletriangle{7} is a newly inserted representative that has been stored after all regular representatives in the vertex buffer. Consequently, the primitive index has to be re-mapped to the corresponding bucketID by means of
\begin{small}
$$
i \mapsto
\begin{cases}
  i - 2 \cdot \text{numBuckets} + 1  & \text{if } i \geq 2 \cdot \text{numBuckets}\\
  i - \text{numBuckets} + 1 & \text{if } i \geq \text{numBuckets}\\
  i & \text{otherwise}\\
\end{cases}     
$$
\end{small}
which is cheap and easy to compute on a GPU.

\section{Handling Updates}
\label{sec:updates}

The presentation thus far has focused on a sorted array-based representation, but inserting into a globally sorted array requires shifting keys and rebuilding the BVH.
Both are prohibitively expensive.
So, to facilitate efficient updates, we propose~\cgrxu{}, a \textit{node-based} variant of this representation.

\begin{figure}[h!]
  \centering
  \includegraphics[page=28, width=\columnwidth, trim={0cm 6.5cm 19.2cm 0}, clip]{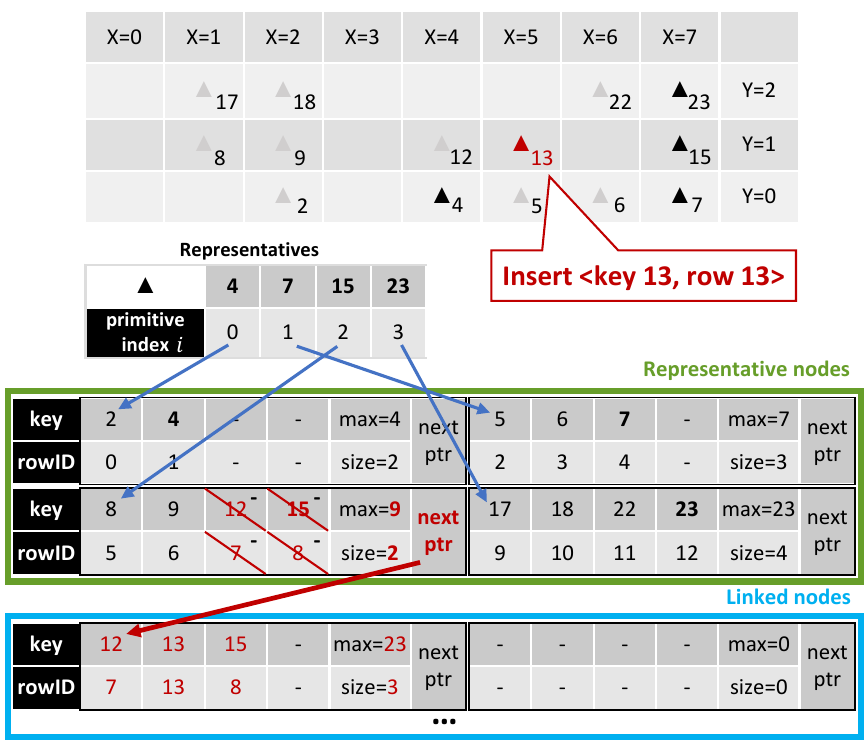}
\vspace*{-0.3cm}    
  \caption{Node-based representation for updates. 
  Memory is partitioned into a green region where the data structure is initially built using contiguous nodes, and a blue region that is used to extend buckets when nodes are split.} 
  \label{fig:updates}
\end{figure}


The high level idea is to implement each \textit{bucket} as a \textit{linked list} of \textit{nodes}.
Nodes have a fixed size~$N$, a tuneable parameter that we analyze in our experiments.
Each node contains sorted \textit{keys} and corresponding \textit{rowID}s, a \textit{next} pointer, a \textit{maxKey} and a current \textit{size}.
For each bucket, an initial \textit{representative node} is created, and subsequent insertions into a bucket will cause this node to be \textit{split}, resulting in the 
creation of a new node, and the movement of half of the keys into that new node.
This is illustrated in Figure~\ref{fig:updates}, which depicts a state some time after the initial construction of the index, after some subsequent insertions have happened, and a key-rowID pair $\langle 13, 13 \rangle$ is inserted.
Nodes in a bucket can be split multiple times, and all nodes corresponding to a bucket are linked together using their \textit{next} pointers, starting with the representative node.
The keys in a bucket are maintained in sorted order across all nodes.
This way, a point lookup terminating at a representative node that has been split can simply follow the \textit{next} pointers to locate the correct node with the relevant key, \textit{without requiring the BVH or buckets to be updated}.

Rather than allocating each node individually, it is more efficient to allocate a large slab of memory, and manually partition it into nodes.
Once this region has been used entirely, we enlarge it by allocating additional memory. 
We divide this large allocation into two subregions, one for representative nodes (\textit{representative node region}), and the other for allocating new nodes to be appended to linked lists (\textit{linked node region}).
Note that the next pointer of nodes can only be directed into the linked node region, since representative nodes are always the heads of their respective linked lists.

\textit{Initial construction.}
Given a sorted array of key-rowID pairs for initial bulk loading, we first divide them into buckets of size $N / 2$ \blue{(half the node size---a tunable parameter).
Note that this partitions the keys in a \textit{distribution-adaptive} way, by evenly dividing the \textit{keys} being bulk loaded across buckets, rather than evenly dividing the \textit{key range}. In other words, every $N/2$-th key in the input array becomes the \textit{maxKey} of a node. A special \textit{overflow bucket} with maxKey $\infty$ is added to handle keys larger than any key in the initial bulk load.}

We then reserve space for one node in the representative node region for each bucket, and fill the node with its corresponding key-rowID pairs in parallel.
Note that each representative triangle effectively points directly to the representative node for its bucket. 
More precisely, since nodes are stored contiguously in the representative node region, the triangle's primitive index can be multiplied by the size of a node, and added to the base address of the representative node region, to obtain the starting address of the representative node.

\textit{Lookups.}
The raytracing procedure to locate a bucket is unchanged, except for the node address calculation explained above.
We then traverse the bucket chain starting at the representative node to find the last node where maxKey is greater than the key we are looking for.
\blue{If there are many duplicates of a single key, these duplicates may span multiple nodes, and they will all be found by this search procedure.}

\textit{Insertion and deletion.}
Similar to lookups, keys to be inserted or deleted are collected in a batch that is then sent to the GPU in an array.
The keys are then sorted.
Any key that is both to be inserted and deleted in a batch can simply be eliminated from the batch.
The actual insertion or deletion is handled by a CUDA kernel that dedicates \textit{one thread to each bucket} in the current representation.
The thread responsible for bucket $i$ does two binary searches on the batch's sorted keys to identify the sequence of keys it is responsible for, and traverses the nodes corresponding to bucket $i$, deleting and inserting these keys as appropriate.
An advantage of allocating one thread per bucket is that there are no concurrency issues associated with updating a bucket (for instance, neither atomic read-modify-write instructions nor locks are needed).

Deletions are processed first, as by doing so, space may be created to facilitate insertions without splitting.
For each key, the thread first locates the appropriate node in the list by comparing with the \textit{maxKey} of each node.
Then, it performs binary search to locate the appropriate index within the appropriate node.
Deletion of a key results in keys to the right being shifted to the left.
Insertion of a key results in shifting to the right.
As explained above, insertion into a full node splits the node, changing its next pointer to point to a new node, and moving half of the keys into the new node.
The new node receives the old node's \textit{maxKey}, and the old node's largest key after the split becomes its new \textit{maxKey}.
If the new node is being inserted in the middle of a list, its next pointer is set to point to the following node.

\section{Parameter Configuration}
\label{sec:configuration}

In the following, we will experimentally analyze the impact of all configuration parameters of~\cgrx{}.
Then, we will use the best configuration(s) in Section~\ref{sec:exp}. 


 
We perform all of the following experiments on an NVIDIA RTX 4090 GPU with 24 GB of VRAM
and 128 raytracing cores. This GPU implements the most recent
Ada Lovelace architecture and is the fastest consumer RTX GPU currently available, as consistently verified during the work on this project. The CPU is an AMD ThreadRipper 3990X.

Unless specified otherwise, we generate a key set of $2^{26}$~keys consisting of 32-bit or 64-bit unsigned integers.
For some fixed integer $d$, the first part of the key set consists of all keys from~$0$ to~\mbox{$d-1$} to reflect a dense key arrangement, and the second part is picked uniformly and randomly from the remaining value range to reflect a sparse key arrangement.
In the experiments, we vary the percentage of keys that are picked uniformly from $0\%$ to $100\%$, which we simply refer to as the \textit{uniformity} of the key set. 
We always shuffle the key sequence, and the final position in the shuffled sequence determines a key's rowID.
Lookups are drawn randomly from the key set, and we perform $2^{27}$~lookups by default.
The rowIDs obtained through the lookup are aggregated per-lookup, and then written to a separate result buffer to test for correctness. 

\subsection{Key Mapping and 3D Representation}
\label{ssec:exp:key_mapping}

First, we compare the naive with the optimized representation of~\cgrx{} in terms of point-lookup performance to see whether one variant performs consistently better than the other.
However, before that, we have to discuss the impact of the key mapping. We identified that for the default key mapping used in~\cite{lit:rtindex}, namely \mbox{$k \mapsto (k_{22:0}, k_{45:23}, k_{63:46})$}, the performance of both variants was not competitive for sparser key distributions. The reason for this is that the proprietary BVH construction algorithm cannot choose a reasonable bounding volume layout due to the data being largely uniform.
This renders the first (and non-avoidable) $x$-axis ray highly expensive.
\begin{figure}[!h]
\centering
\begin{subfigure}[b]{0.47\columnwidth}
\centering
\includegraphics[page=3, width=.8\columnwidth, trim={0 19.7cm 29.5cm 1.7cm}, clip]{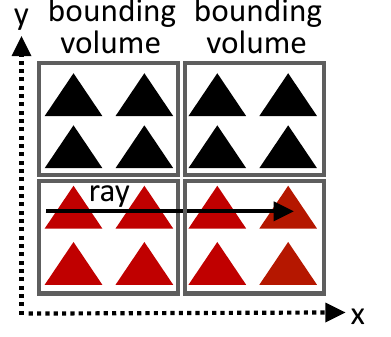}
\caption{No scaling.}
\label{fig:clustering:no_scaling}
\end{subfigure}
\hspace*{0.2cm}
\begin{subfigure}[b]{0.47\columnwidth}
\centering
\includegraphics[page=4, width=0.8\columnwidth, trim={0 19.7cm 29.5cm 1.7cm}, clip]{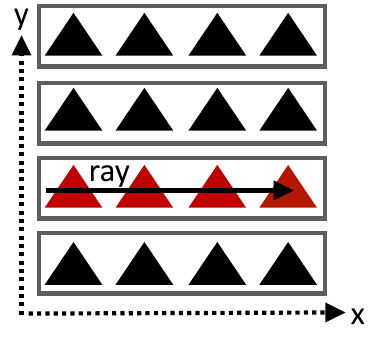}
\caption{Scaling the $y$-axis.}
\label{fig:clustering:scaling}
\end{subfigure}
\caption{The impact of scaling on the BVH structure (shown for 2D). Red triangles must be tested for intersection.}
\label{fig:clustering}
\end{figure}
In Figure~\ref{fig:clustering:no_scaling}, we conceptually visualize such a disadvantageous clustering.
In the example, the $x$-axis ray has to perform costly intersection tests with an unnecessarily large number of triangles since it also has to check for intersections in neighboring rows. 
Ideally, the bounding volumes would primarily extend along the $x$-axis, such that only the triangles in the current row have to be checked for intersection, as shown in Figure~\ref{fig:clustering:scaling}.
To incentivize such a grouping, we simply adjust the key mapping slightly by multiplying both the $y$-coordinate~$k_{45:23}$ and the $z$-coordinate~$k_{63:46}$ with a large, carefully chosen constant, resulting in the mapping \mbox{$k \mapsto (k_{22:0}, 2^{15} \cdot k_{45:23}, 2^{25} \cdot k_{63:46})$}. 
\textit{Consequently, in all upcoming experiments, we only use the scaled mapping.} 

\begin{figure}[h!]
    \centering
\includegraphics[width=\columnwidth, trim={0 0.4cm 0 0}, clip]{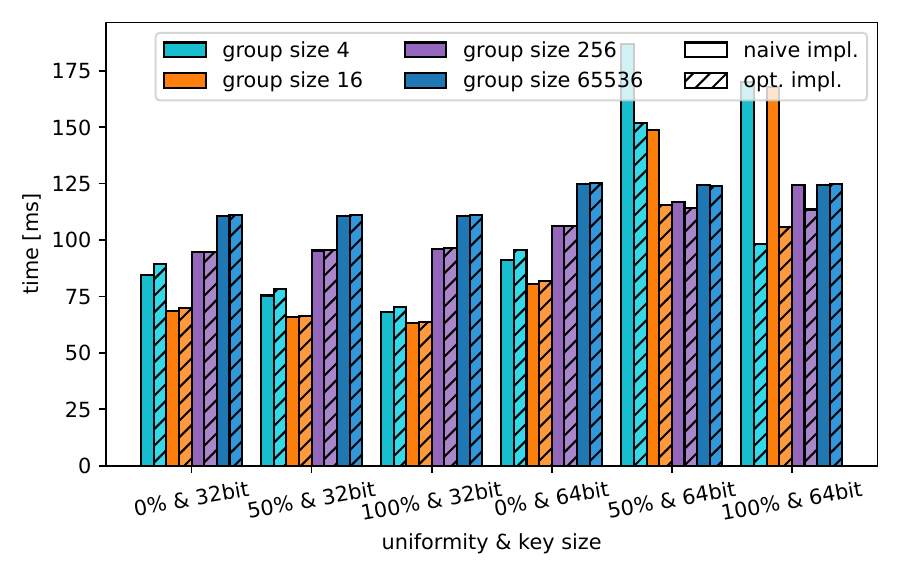}
\vspace*{-0.3cm}
    \caption{Naive vs optimized representation for scaled key mapping $k \mapsto (k_{22:0}, 2^{15} \cdot k_{45:23}, 2^{25} \cdot k_{63:46})$.}
    \label{fig:exp:key_mapping_all_distributions:scaled}
\end{figure}



Coming back to the initial comparison of the naive and optimized representations, we can observe in Figure~\ref{fig:exp:key_mapping_all_distributions:scaled} that for 32-bit key sets, both variants perform equally well.
From a geometric perspective, 32-bit keys are always arranged on a single plane.
The amount of rays for lookups is therefore limited to three, with most lookups requiring only one ray.
Therefore, the optimized scene representation does not yield significant improvements.
However, for the 64-bit key sets with a high uniformity (and hence, high sparsity), optimizing the representatives significantly shortens the lookup procedure and improves the performance.
Inspecting the number of individually fired rays for each variant revealed that for smaller buckets, the optimized representation avoids firing the second $x$-axis ray in most cases, because the previous $y$-axis ray hit a flipped representative.
Apart from performance, we can also confirm that the optimized representation reduces the memory footprint over the naive representation for sparse key sets:
For example, for 64-bit keys and a bucket size of 4, the optimized representation saves $16\%$ and $28\%$ memory over the naive one for a uniformity of $50\%$ and $100\%$, respectively.
\textit{Hence, we will use only the optimized representation in the following.}

\subsection{Bucket Size}
\label{ssec:bucket_size}

\blue{Next, we identify the best bucket size for \cgrx{}. As our index aims at providing high space efficiency while achieving competitive lookup performance, intuitively, we want to choose the largest bucket size that still offers a good throughput. To quantify how well our index balances this trade-off, we introduce a metric called \textit{throughput per memory footprint}. We take the throughput as entries looked up per second and divide it by the memory footprint of the structure in bytes. By this, we essentially measure how an index structure ``buys'' throughput performance by consuming additional memory (for \cgrx{}, this means adding representatives in the scene).}

\begin{figure}[h!]
  \centering
  \begin{subfigure}[b]{\linewidth}
    \includegraphics[page=2, width=\linewidth, trim={0 7.2cm 0 0}, clip]{plots/partition-size-heatmaps/4090/heatmap-32-26.pdf}
    \caption{Point-lookup time.}
    \label{fig:exp:heatmap:time}
  \end{subfigure}
  \begin{subfigure}[b]{\linewidth}
    \includegraphics[page=1, width=\linewidth, trim={0 7.2cm 0 0}, clip]{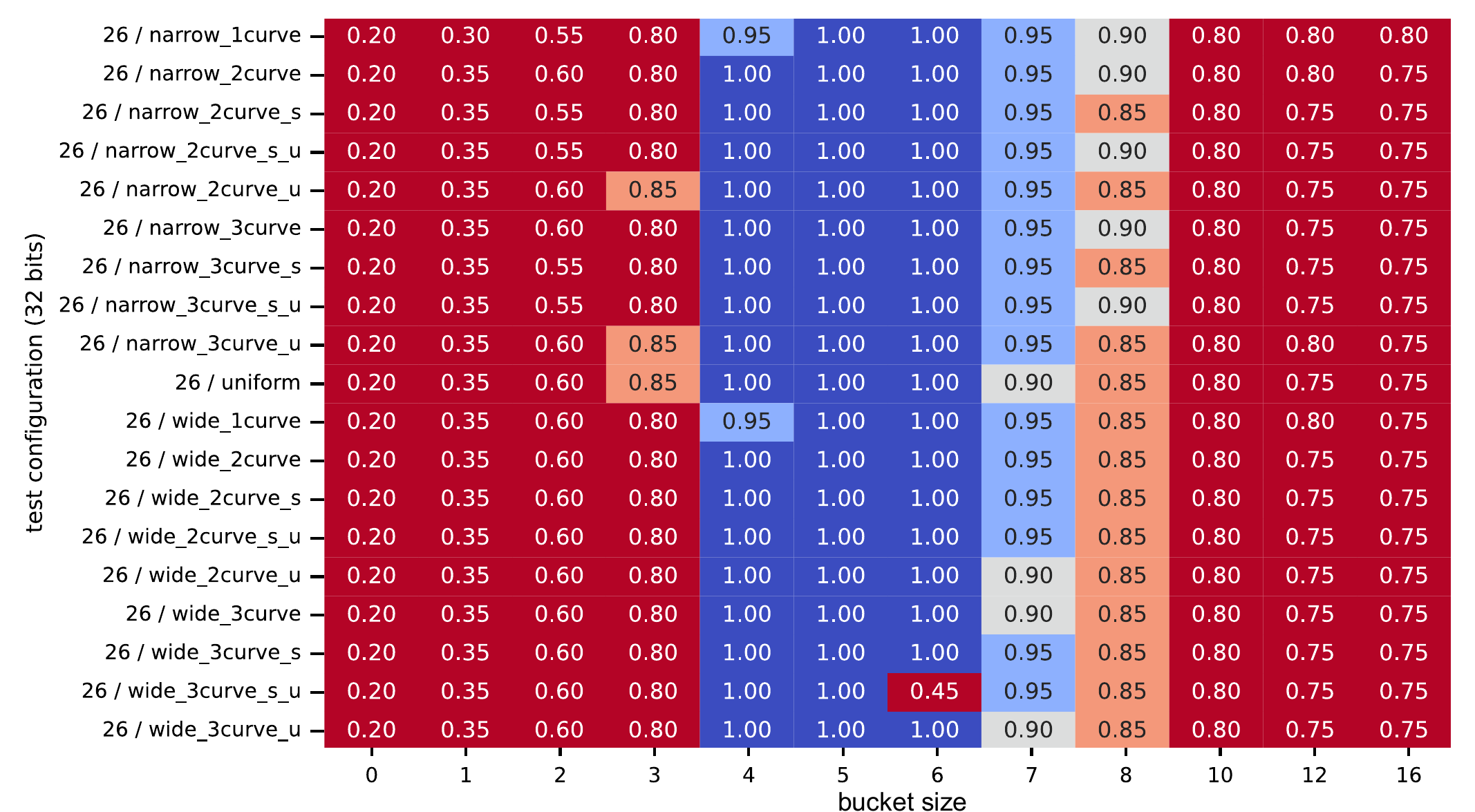}
    \caption{Throughput per memory footprint.}
    \label{fig:exp:heatmap:tp_per_footprint}
  \end{subfigure}
  \caption{\blue{Robustness of bucket size across 19 key distributions.}}
    \label{fig:exp:heatmap}
\end{figure}

\begin{figure*}[!h]
\centering
\begin{subfigure}[b]{0.3\textwidth}
\includegraphics[width=1.0\linewidth, trim={0.4cm 0 0 0}, clip]{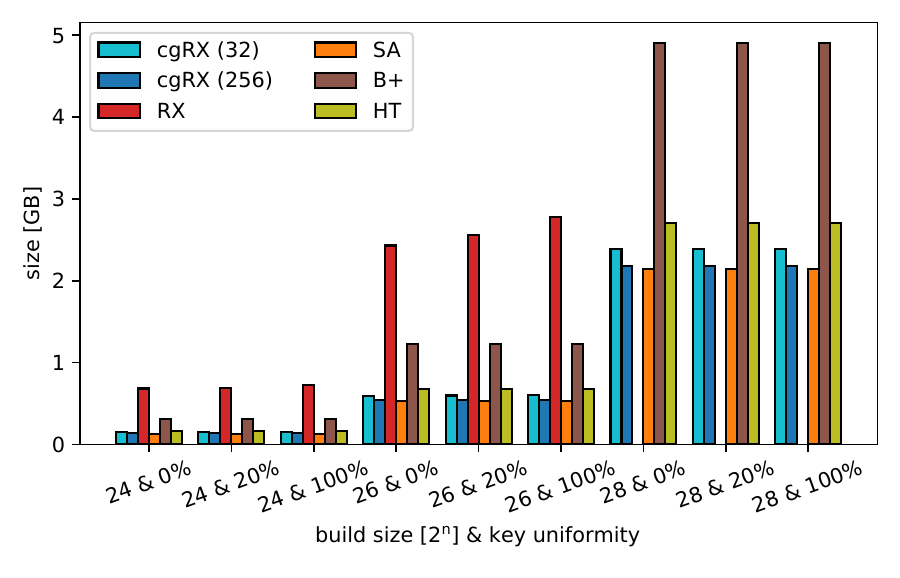}
\caption{Memory footprint.}
\label{fig:exp:size:32}
\end{subfigure}
\begin{subfigure}[b]{0.3\textwidth}
\includegraphics[width=1.0\linewidth, trim={0.4cm 0 0 0}, clip]{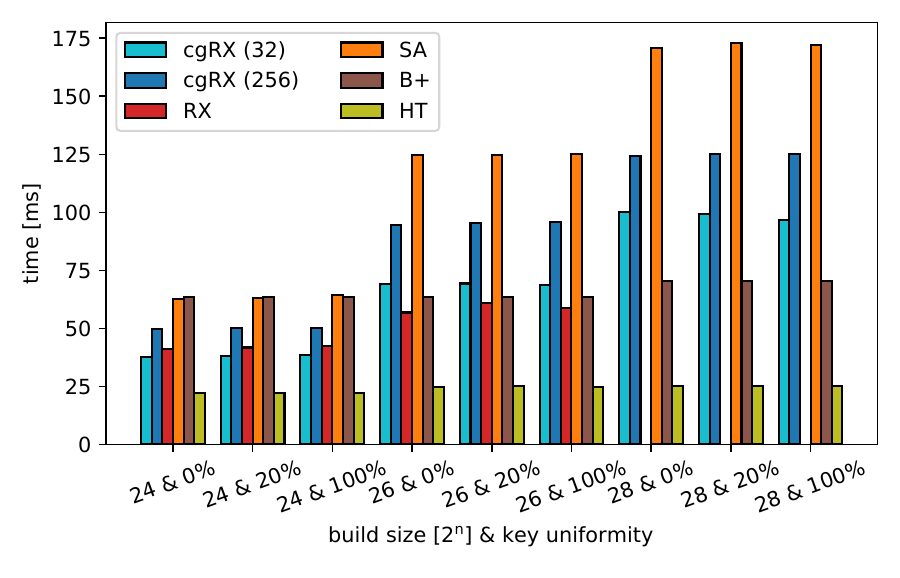}
\caption{Accumulated point-lookup time.}
\label{fig:exp:point_lookups:32}
\end{subfigure}
\begin{subfigure}[b]{0.3\textwidth}
\includegraphics[width=1.0\linewidth, trim={0.4cm 0 0 0}, clip]{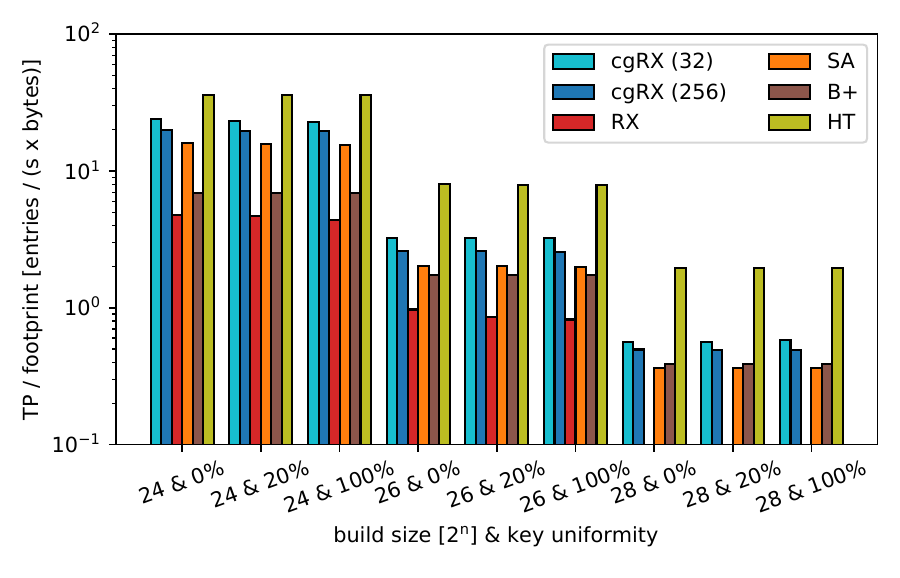}
\caption{Throughput per memory footprint.}
\label{fig:exp:point_lookups_per_footprint:32}
\end{subfigure}
%
\caption{Comparison of memory footprint and point-lookup performance for key range $[0, 2^{32}-1]$.}
\label{fig:exp:memory_footpring_and_point_lookups:32}
\end{figure*}

\begin{figure*}[!h]
\centering
\begin{subfigure}[b]{0.3\textwidth}
\includegraphics[width=1.0\linewidth, trim={0.4cm 0 0 0}, clip]{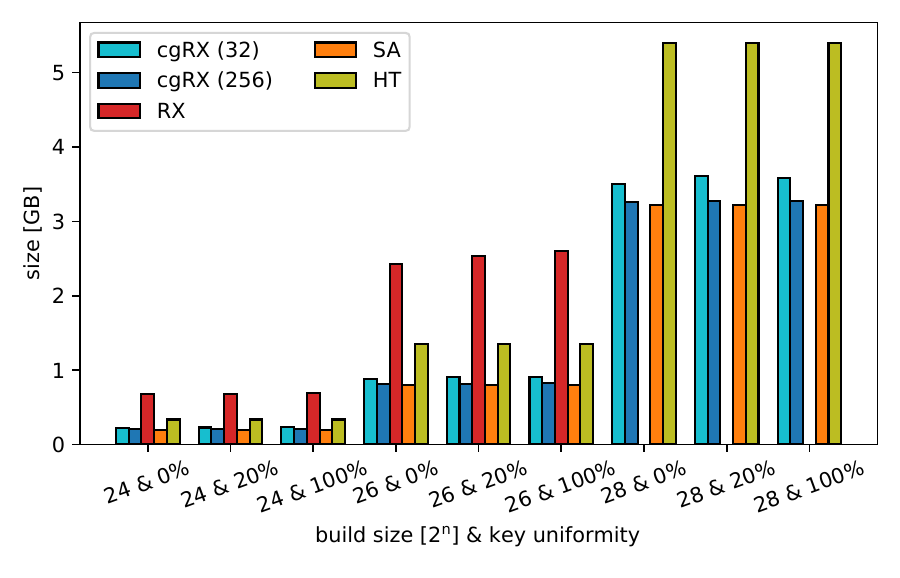}
\caption{\blue{Memory footprint.}}
\label{fig:exp:size:64}
\end{subfigure}
\begin{subfigure}[b]{0.3\textwidth}
\includegraphics[width=1.0\linewidth, trim={0.4cm 0 0 0}, clip]{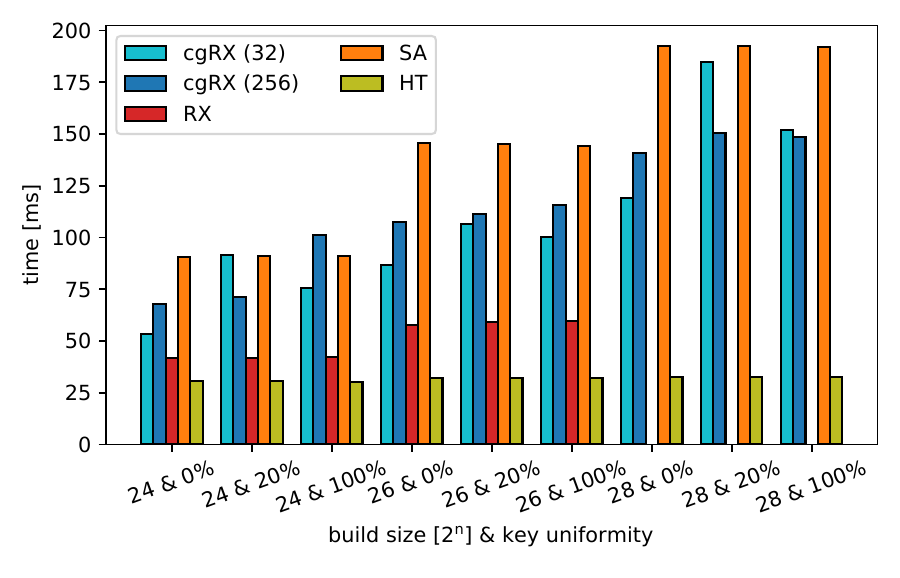}
\caption{\blue{Accumulated point-lookup time.}}
\label{fig:exp:point_lookups:64}
\end{subfigure}
\begin{subfigure}[b]{0.3\textwidth}
\includegraphics[width=1.0\linewidth, trim={0.4cm 0 0 0}, clip]{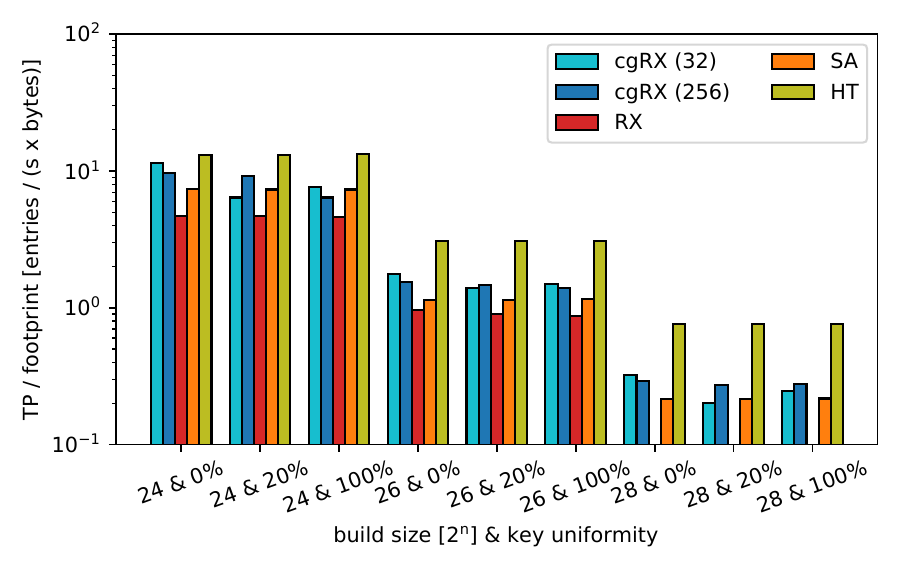}
\caption{\blue{Throughput per memory footprint.}}
\label{fig:exp:point_lookups_per_footprint:64}
\end{subfigure}
%
\caption{\blue{Comparison of memory footprint and point-lookup performance for key range $[0, 2^{64}-1]$.}}
\label{fig:exp:memory_footpring_and_point_lookups:64}
\end{figure*}

\blue{To analyze which bucket size performs best, we evaluate twelve bucket size configurations against nineteen different key distributions, varying from uniform to highly skewed and mixtures of both. Further, we test both 32-bit and 64-bit keys, five different key set sizes ($2^{24}$ to $2^{28}$), and two GPU generations (RTX 4090 and A6000), resulting in $4560$ test combinations in total. 
The results show that across the board, a universal decision on the best performing bucket size can be made.  
Due to space restrictions, in Figure~\ref{fig:exp:heatmap}, we can show only the results for 32-bit keys and a key set size of $2^{26}$ on the RTX~4090. For each configuration, we show the relative performance in comparison to the best performing bucket size, where Figure~\ref{fig:exp:heatmap:time} shows the pure point lookup-time, while Figure~\ref{fig:exp:heatmap:tp_per_footprint} shows the throughput per memory footprint metric.
We can see that for the latter, a bucket size of $2^5=32$ performs best in the vast majority of cases. Even when inspecting only the lookup performance, a bucket size of $32$ is among the best choices. However, we can also see that for a significantly larger bucket size of $2^8=256$, the throughput per memory footprint is still $85\%$ of the best option. Consequently, we also consider this configuration further as a space efficient alternative.}

\blue{\textit{Size recommendation}.
Based on our findings, \cgrx{} uses a bucket size of \textbf{32} by default for the remaining experiments, which optimizes the throughput per memory footprint ratio. Also, we show the results for a bucket size of \textbf{256} as a space efficient and performance-competitive alternative.}

\vspace*{-0.1cm}
\section{Experimental Evaluation}
\label{sec:exp}
\vspace*{-0.1cm}

In the following, we evaluate how assorted configurations of~\cgrx{} perform against a set of competitive baselines. Under the update workload, we evaluate both \cgrx{} and \mbox{\cgrxu{}}.
We include the same baselines as in~\cite{lit:rtindex}: \hash{}: A GPU-resident open addressing hash table~\cite{lit:warpcore, lit:hash-warpcore}, which performs cooperative probing. The target load factor is set to the recommended~$80\%$ ($40\%$ for updates).
\bp{}: A GPU-resident B+tree~\cite{lit:btree2, lit:btree-code} performing cooperative 16-thread tree traversal and only supporting 32-bit keys.
\sa{}: A GPU-resident Sorted Array~\cite{lit:rtindex} which uses binary search for lookups. \blue{Note that all methods that require the sorting of the input key-rowID array (\cgrx{}, \bp{}, and \sa{}) use CUB's \smalltt{DeviceRadixSort}~\cite{lit:cub} for this purpose, and the \textbf{cost for sorting is always included} in the reported times.}

\begin{table}[h!]
    \footnotesize
    \hspace*{-0.25cm}
    \centering
    \begin{tabular}{C{1.53cm}|c|c|c|c|c|c}
        Method & Point & Range & Mem & 64-bit & Bulk-load & Updates\\\hline
        \textbf{HT} \cite{lit:warpcore, lit:hash-warpcore} & \checkmark & $\times$ & med & \checkmark & $\times$ & \checkmark \\\hline
        \textbf{B+} \cite{lit:btree2, lit:btree-code} & \checkmark & \checkmark & med & $\times$ & \checkmark & \checkmark \\\hline
        \textbf{SA} \cite{lit:rtindex} & \checkmark & \checkmark & low &\checkmark & \checkmark & rebuild \\\hline 
        \textbf{RX} \cite{lit:rtindex} & \checkmark & \checkmark & high & \checkmark & \checkmark & rebuild \\\hline
        \textbf{RTScan (RTc1)} \cite{lit:rtscan} & $\times$ & \checkmark & high & limited & on CPU & rebuild \\\hline
        \textbf{cgRX} & \checkmark & \checkmark & low & \checkmark & \checkmark & rebuild \\\hline
        \textbf{cgRX-u} & \checkmark & \checkmark & low & \checkmark & \checkmark & \checkmark 
    \end{tabular}
    \caption{Overview of all tested indexes.}
    \label{tab:comparison}
\end{table}

Of course, we also compare against the original fine-granular~\textbf{RX}. 
Additionally, we compare the range-lookup performance against \blue{\rtscan{}~\cite{lit:rtscan}}, another recently published raytracing based indexing method which has been specifically designed for that purpose. Instead of concurrently executing a large number of (single-threaded) lookups, \rtscan{} parallelizes a single range lookup by firing a large number of rays at different positions concurrently, where the number of concurrently fired rays depends on the size of the range. To ensure a fair comparison for the case where a single range lookup does not fully utilize the available resources, we extended the original implementation to support executing a batch of $32$~range lookups concurrently. Note that \rtscan{} does not support point lookups out of the box and hence, we cannot include it in the corresponding experiments. Table~\ref{tab:comparison} summarizes the core features of all competitors. 
 
\subsection{Memory Footprint and Point-lookup Performance}

One of our central motivations for \cgrx{} was to reduce the memory footprint of the original~\rx{} approach \textit{while} providing good performance.
To find out whether \cgrx{} achieves this goal, Figure~\ref{fig:exp:size:32} shows the permanent memory footprint of all methods on 32-bit key sets of different sizes (\blue{$2^{24}$}, $2^{26}$, and $2^{28}$) with varying uniformity ($0\%$, $20\%$, and $100\%$). For \cgrx{}, we evaluate \blue{bucket sizes~$32$ and $256$}.
While for \blue{$2^{24}$}~keys, all indexes have a negligible memory footprint, for $2^{26}$~keys, \rx{} has by far the highest footprint between $2.5$GiB ($0\%$ uniformity) and $2.9$GiB ($100\%$ uniformity).
In comparison, \blue{even for a rather small bucket size of~$32$,~\cgrx{} shows a significantly lower memory footprint of only around $0.7$GiB}, which is already less than the footprint of~\bp{} at around~$1.1$GiB.
\blue{For a bucket size of~$256$, \cgrx{} even approaches the space-optimal \sa{}.}
For $2^{28}$~keys, \rx{} runs out of memory, \blue{while \cgrx{} still beats \hash{} and stays on par with \sa{}}. \bp{} consumes almost twice the amount of memory as the best \cgrx{} configuration. 

In the Figures~\ref{fig:exp:point_lookups:32} and \ref{fig:exp:point_lookups_per_footprint:32}, we bring the point-lookup performance into the picture. While Figure~\ref{fig:exp:point_lookups:32} showing the accumulated point-lookup time alone reveals that \cgrx{} has a point-lookup performance somewhere between the remaining range-lookup supporting structures, Figure~\ref{fig:exp:point_lookups_per_footprint:32} sets both memory footprint and point-lookup performance into perspective \blue{by showing the previously introduced throughput per memory footprint metric}. 
We can see that  \cgrx{} clearly performs best among all indexes supporting range lookups.
For \blue{$2^{24}$~keys}, the best configuration of \cgrx{} has a throughput per memory footprint that is $5\times$ higher than for \rx{}, $3.5\times$ higher than for \bp{}, and $1.5\times$ higher than for \sa{}.
At $2^{26}$~keys, the throughput per memory footprint is still $3.4\times$ higher than for \rx{}, $1.9\times$ higher than for \bp{}, and $1.6\times$ higher than for \sa{}. Only \hash{} outperforms \cgrx{} by $2.5\times$.  
For $2^{28}$~keys, where \rx{} runs out of memory, the improvement over \bp{} and \sa{} is still around $1.5\times$. 
\blue{Note that we observe a similar trend for 64-bit keys in Figure~\ref{fig:exp:memory_footpring_and_point_lookups:64} where, unfortunately, we cannot include \bp{} as it lacks the support for wide keys.}
Overall, we can see that \cgrx{} provides the best ``bang for the buck'' of all general-purpose GPU-resident indexes, a property which is especially important in the presence of scarce GPU memory. 

\subsection{Range-lookup Performance}

Next, we analyze the range-lookup performance. 
We use a 32-bit key set of $2^{26}$ keys with a uniformity of $0\%$ (dense) and vary the number of expected hits per range lookup from $2^0$, which resembles a point lookup, to $2^{24}$. We fire a batch of $2^{16}$ range lookups and  report the normalized cumulative lookup time, which is the total time of all range lookups of the batch divided by the total number of retrieved entries, on a logarithmic scale. 
We also include the range-lookup specific baselines \rtscan{} as well as a \scan{} which just scans the entire array and filters for the range of interest. \hash{} does not support range lookups and is therefore not included. 

\begin{figure}[h!]
    \centering
\includegraphics[width=0.85\linewidth]{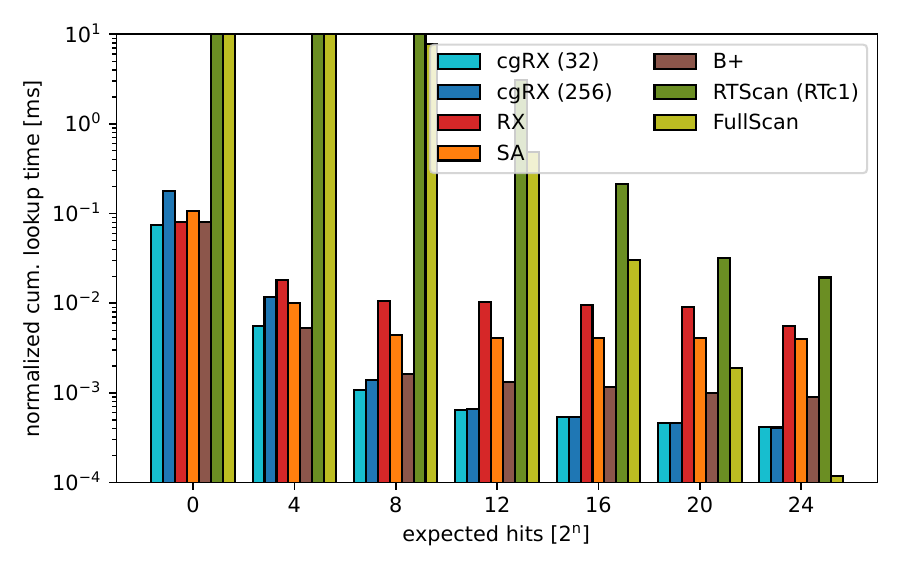}
    \caption{Range lookups on a dense 32-bit key range.}
    \label{fig:exp:range_lookups}
    \vspace*{-0.3cm}
\end{figure}

Figure~\ref{fig:exp:range_lookups} shows the results (cut off at 10ms).
We can see that for range lookups, \cgrx{}~(32) outperforms the direct competitor \rx{} for all tested bucket sizes.
The reason for this is that \cgrx{} must perform only one point lookup per range lookup, followed by a simple scan.
In contrast, \rx{} must detect all qualifying entries in the collision detection of its tracing procedure, which is prohibitively expensive.  
\blue{For selectivities between $2^8$ and $2^{20}$~hits per range lookup, \cgrx{} is the best performing method. For example, for $2^8$, it is more than $9.9\times$ faster than \rx{}, $4.2\times$ faster than \sa{}, and $1.5\times$ faster than \bp{}.}
\blue{As \cgrx{} uses a cooperative scan just like \bp{}, the performance of both methods becomes almost the same for lower selectivities. However, \cgrx{} has the advantage that all data is stored consecutively, while \bp{} scans individual leaf nodes.} Surprisingly, \rtscan{} is even four orders of magnitude slower than \cgrx{}, as it is not able to properly parallelize the batch of lookups. In fact, \rtscan{} is even slower than \scan{}, showing that it is currently not suited for answering batched range lookups, despite batched lookups being a common workload on GPUs.

\begin{figure}[h!]
    \centering
\includegraphics[width=0.85\linewidth]{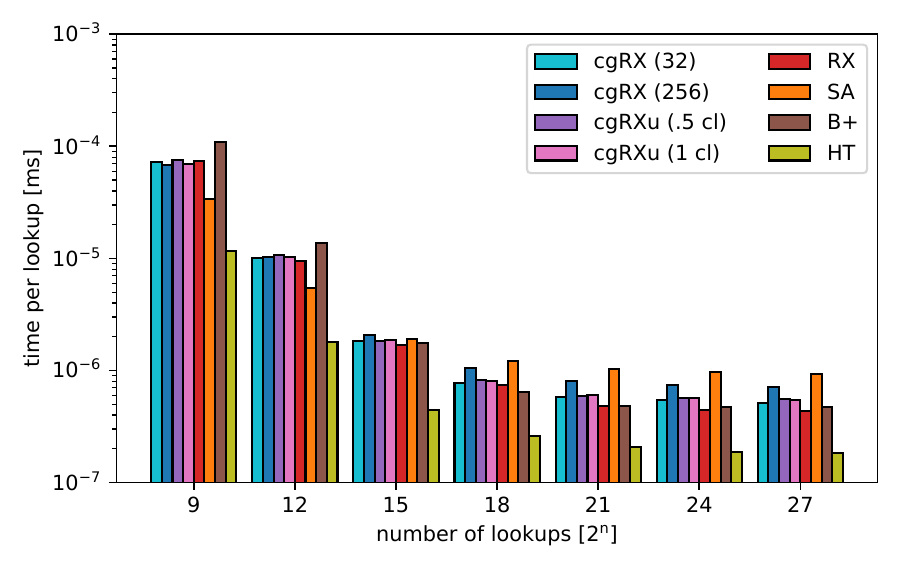}
\vspace*{-0.3cm}
    \caption{Varying the number of lookups fired in a batch.}
    \label{fig:exp:probe_scaling}
    \vspace*{-0.3cm}
\end{figure}

\subsection{Varying the Batch Size}

So far, we have used a batch size of~$2^{27}$ point lookups and $2^{16}$ range lookups. Depending on the application, smaller or larger batches might also arrive, thus we now vary the number of point lookups fired in a batch from $2^{9}$ to $2^{27}$. \blue{For completeness, we also include \cgrxu{} here.} Figure~\ref{fig:exp:probe_scaling} reports for each configuration the time per lookup, i.e., the time it took to answer all queries divided by the number of queries, on a logarithmic scale. 
First of all, we can observe that the performance of all methods deteriorates with a decrease in batch size --- for very small batches of $2^{9}$ and $2^{12}$~point lookups, the GPU becomes severely under-utilized. For batch sizes of $2^{15}$ to $2^{27}$, the performance remains rather stable for all indexes. This shows that \cgrx{} is not more susceptible to smaller batch sizes than the traditional baselines --- in contrast, for a batch size of $2^{15}$, \cgrx{} draws even with \rx{} and \bp{} while having a significantly smaller memory footprint. 

\begin{figure}[h!]
    \centering
\includegraphics[width=.85\linewidth]{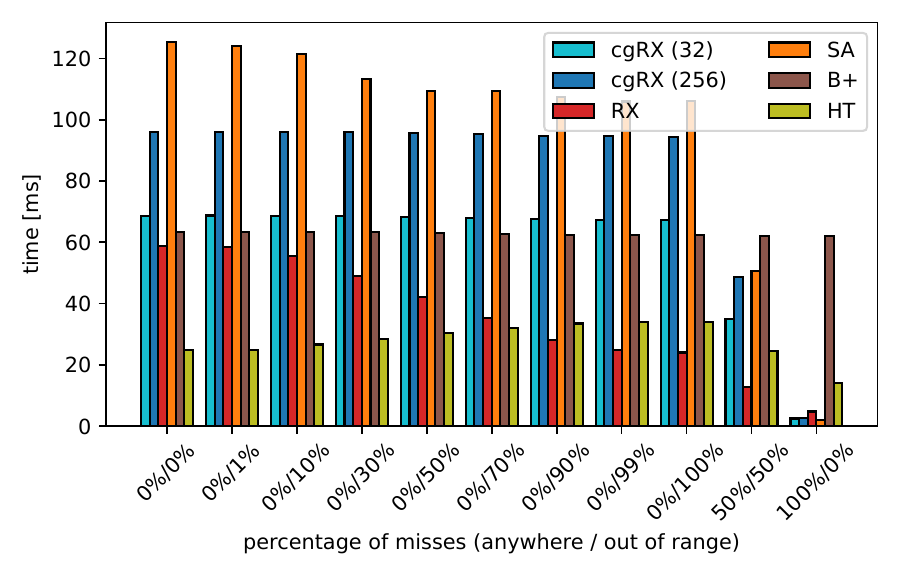}
\vspace*{-0.3cm}
    \caption{Varying the hit ratio.}
    \vspace*{-0.3cm}
    \label{fig:exp:hit_ratio}
    \vspace*{-0.2cm}
\end{figure}

\subsection{Varying the Hit Ratio}

So far, all point lookups resulted in hits.
To see the impact of misses, we now fire a certain amount of point lookups that do \textit{not} hit an indexed key and report the accumulated point-lookup time.
We differentiate between misses that lie within the value range of the indexed data, and ones that lie outside of that range.
The key set consists of 32-bit keys with uniformity~$100\%$.  
Figure~\ref{fig:exp:hit_ratio} shows the results.
\begin{figure}[h!]
    \centering
\includegraphics[width=.85\linewidth]{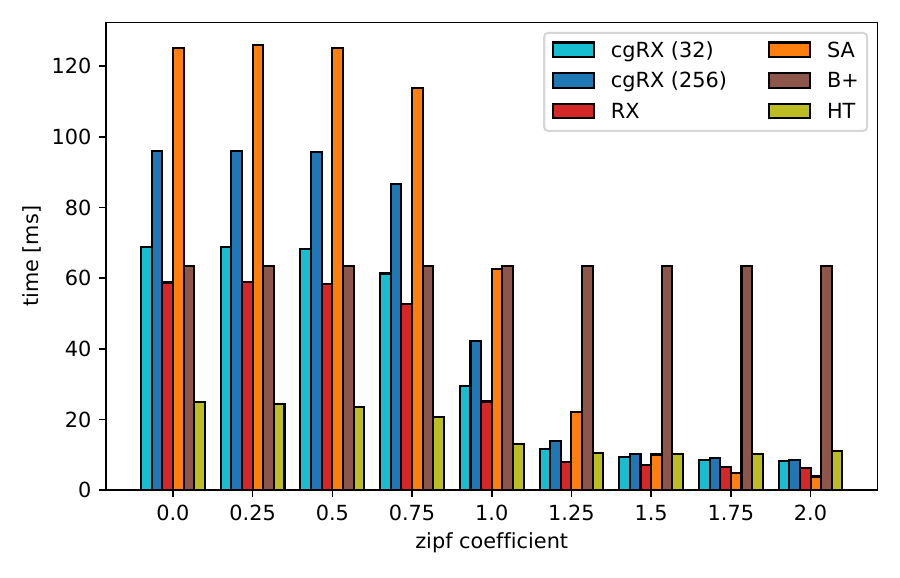}
\vspace*{-0.3cm}
    \caption{Varying the skew of lookups.}
    \label{fig:exp:skewed_lookups}
\end{figure}
\begin{figure*}[!h]
\centering
\begin{subfigure}[b]{0.32\textwidth}
\includegraphics[width=1.0\linewidth, trim={0.4cm 0 0 0}, clip]{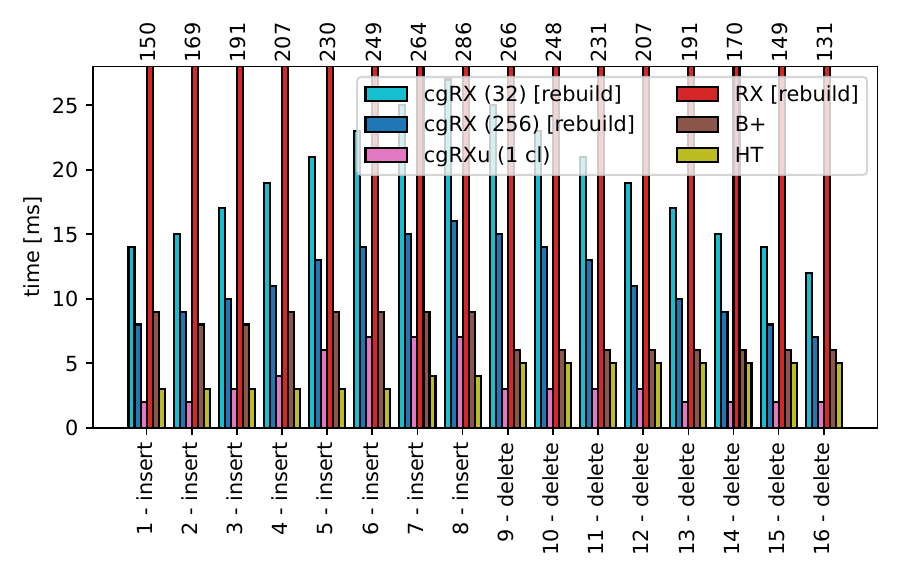}
\caption{Applying batches of updates.}
\label{fig:exp:updates:update_time}
\end{subfigure}
\begin{subfigure}[b]{0.32\textwidth}
\includegraphics[width=1.0\linewidth, trim={0.4cm 0 0 0}, clip]{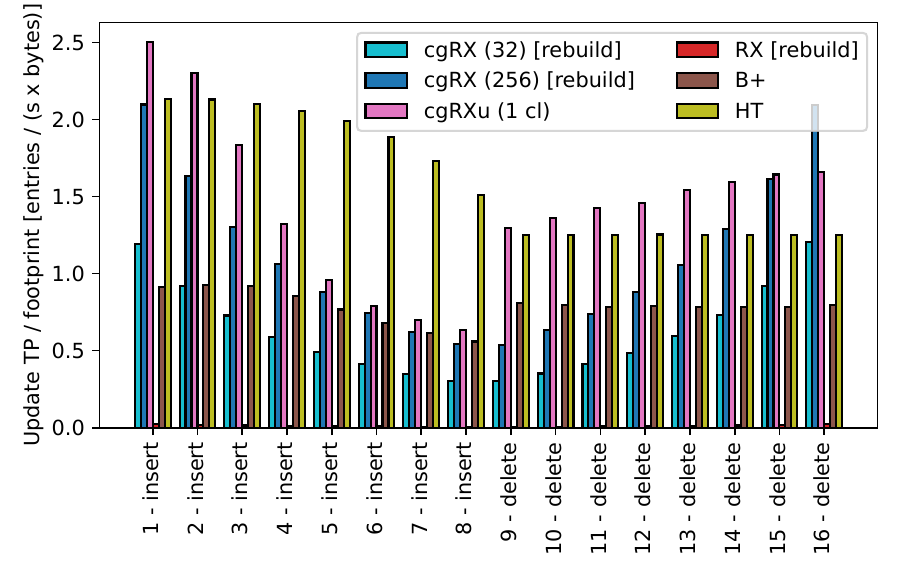}
\caption{Update throughput per mem. footprint.}
\label{fig:exp:updates:tp_per_footprint}
\end{subfigure}
\begin{subfigure}[b]{0.32\textwidth}
\includegraphics[width=1.0\linewidth, trim={0.4cm 0 0 0}, clip]{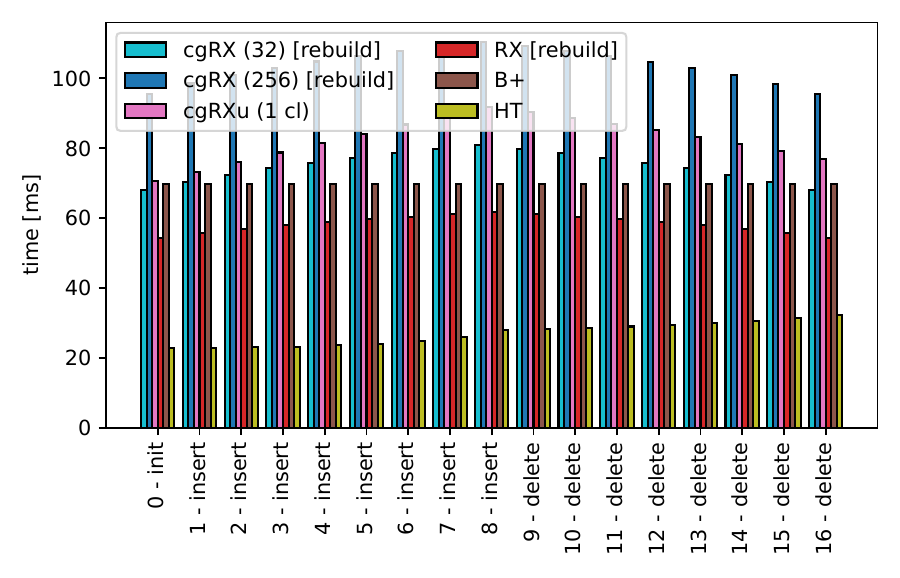}
\caption{Accumulated point-lookup time.}
\label{fig:exp:updates:lookup_time}
\end{subfigure}
\caption{Updating \cgrxu{}, \bp{}, and \hash{} compared against rebuilding \cgrx{} and \rx{} from scratch.}
\label{fig:exp:updates}
\end{figure*}
We observe that while \rx{} strongly benefits from misses, this is not the case for \cgrx{}.
The reason for this is that \rx{} is able to abort the BVH traversal as soon as it detects that a key is not covered by any bounding volume.
This is not possible for \cgrx{} which always finds a representative if the target key is within the value range. 
Consequently, the miss is detected rather late during the bucket search process.
This means that \cgrx{} should be primarily used in hit-only or hit-mostly lookup scenarios. 
The last two bars show the effect of out-of-range misses:
Here, the search is trivial, and \cgrx{} can answer all queries quickly.

\subsection{Varying the Lookup Skew}

In all previous experiments, we picked the lookup keys uniformly from the key range.
In the following, we test the effect of skewed lookups, which follow a Zipf distribution.
We include the uniform distribution seen so far by using a Zipf coefficient of $0.0$, and evaluate different levels of skew by varying the coefficient from $0.25$~(low skew) to $5.0$~(extreme skew).
Again, we report the accumulated point-lookup time.

In Figure~\ref{fig:exp:skewed_lookups}, we can observe that skew is generally beneficial from a performance perspective, as it increases the chance of cache hits and therefore reduces memory accesses.
\bp{} is an outlier here, where the lookup time is apparently unaffected by skew.
NVIDIA's kernel profiler shows that the execution is bottlenecked by the so-called address divergence unit, which handles block synchronization and divergent branches.

\vspace*{-0.1cm}
\subsection{Updates}
\label{ssec:exp:updates}
\vspace*{-0.1cm}

Finally, let us investigate how well our node-based update mechanism in \cgrxu{} performs. \blue{We compare it against the alternative of rebuilding \cgrx{} from scratch for every update batch. Further, we include the option of rebuilding \rx{}, which was \blue{also} the only practical way of applying updates in~\cite{lit:rtindex}, as well as the update mechanisms of \bp{} and \hash{}}.

Initially, we bulk-load all variants with $2^{26}$ keys, with $100\%$ uniformity. Then, we fire eight waves of equally-sized insertion batches, where each batch is followed by a lookup batch of size $2^{27}$. We configure these waves of insertions such that in total, the number of entries is increased by $2.2\times$.
Afterwards, we perform eight corresponding waves of deletions, again interleaved with lookups.
We configure \cgrxu{} with a \blue{node size corresponding to a 128B cache line, initially filled to 50\%}.

Figure~\ref{fig:exp:updates:update_time} shows the time to apply update waves for all variants, while Figure~\ref{fig:exp:updates:tp_per_footprint} shows the ratio of update throughput to the structure's current memory footprint. Figure~\ref{fig:exp:updates:lookup_time} shows the time to perform lookup batches after each update wave has been applied. 
\blue{We can see in~\ref{fig:exp:updates:update_time} that \cgrxu{} reduces the cost of applying updates significantly by up to \blue{$5.6\times$} in comparison to the impractical full rebuilds in \cgrx{}. 
Compared to other baselines (\bp{} and \hash{}), \cgrxu{} indicates competitive update performance and even outperforms most baselines in many instances.
Also, the cost of updating \cgrxu{} increases at a slower rate than the cost of fully rebuilding. Rebuilding \rx{} is not an option at all, as it is an order of magnitude more expensive than the baselines. At the same time, we can observe in~\ref{fig:exp:updates:tp_per_footprint} that introducing a linked list of nodes to represent buckets in \cgrxu{} results in improved update throughput per memory footprint, even though nodes might only be partially occupied.} 
However, we can also see in~\ref{fig:exp:updates:lookup_time} that for update-heavy workloads \bp{} and \hash{} still perform lookups better than \cgrx{} and \cgrxu{}. 

\vspace*{-0.0cm}
\section{Related Work}
\label{sec:related_work}
\vspace*{-0.0cm}

Apart from the original \rx{}~\cite{lit:rtindex} and the recently published \rtscan{}~\cite{lit:rtscan}, which are both baselines in this work, there exists a line of work from other areas that exploits hardware-accelerated raytracing.
These include point containment tests~\cite{lit:app-containment-1, lit:app-containment-2, lit:app-containment-3, lit:app-containment-4}, time-of-flight imaging~\cite{lit:app-tof-1, lit:app-tof-2, lit:app-tof-3}, radius or nearest neighbour search~\cite{lit:app-neighbor-1, lit:app-neighbor-2,lit:app-arkade}, graph rendering ~\cite{lit:app-graph} and tracking of particle movement in applied physics~\cite{lit:app-physics-1, lit:app-physics-2, lit:app-physics-3}. \textcolor{black}{These applications re-write their original problem into a ray tracing task, then leverage RT-core hardware acceleration for efficient processing. RayJoin~\cite{lit:app-rayjoin} is a recent application leveraging RT cores to accelerate spatial join queries. In contrast to traditional relational joins, these spatial joins require fast evaluation of line segment intersections and point-in-polygon tests.
Arkade~\cite{lit:app-arkade} builds on prior research in k-nearest neighbor search using ray tracing, expanding support to include additional distance metrics such as Manhattan and angular (cosine) distances.
While these works also exploit hardware-accelerated ray tracing creatively, they unfortunately do not qualify as indexing baselines for this work.}
In terms of indexes, many CPU data structures have been adapted to become GPU-resident in recent years.
GPU-resident indexes include hash tables~\cite{lit:hash-warpdrive, lit:hash-warpcore, lit:hash-slabhash, lit:hash-megakv, lit:hash-cudpp1, lit:hash-cudpp2, lit:hash-dycuckoo}, from which we picked our baseline \hash{}, but also bloom filters~\cite{lit:bloomfilter1, lit:bloomfilter2, lit:hash-warpcore} and quotient filters~\cite{lit:quotientfilter}, which are suitable for set containment tests and trade memory footprint with false-positive accuracy. Further, radix trees~\cite{lit:radixtree} and comparison-based trees~\cite{lit:fasttree, lit:lsmtree, lit:btree1, lit:btree2} also exist for GPUs and also provide range lookup support. While our evaluation includes a state-of-the-art comparison-based tree \bp{}, unfortunately, no public implementation of the radix trees is available at the time of writing.
There also exist GPU-resident spatial indexes such as R-Trees~\cite{lit:rtree1, lit:rtree2}, GPU permutation indexes~\cite{lit:permutationindex}, and a GPU-resident learned index~\cite{lit:learned_index_GPU}. While these would also make great baselines for our comparisons, their codebase is not publicly accessible.


\vspace*{-0.1cm}
\section{Conclusion}
\label{sec:conclusion}
\vspace*{-0.0cm}

We presented~\cgrx{}, a coarse-granular GPU-resident index which exploits hardware acceleration via RT cores and overcomes the main limitations of its fine-granular predecessor~\rx{}, namely high memory footprint, poor range-lookup performance, and bad updateability. We have shown that \cgrx{} provides the most bang for the buck by offering an up to $5\times$ higher throughput in relation to the memory footprint compared to other state-of-the-art GPU-resident indexes that support both point and range lookups. At the same time, \cgrx{}~ improves the range-lookup performance more than $15\times$ over \rx{}. Our updatable variant \cgrxu{} improves the update performance by up to $5.6\times$ over rebuilding from scratch. 

\bibliographystyle{IEEEtran}
\bibliography{cgRTIndeX}

\end{document}